\documentclass[twocolumn]{aa}
\usepackage{longtable}
\usepackage{multirow}
\usepackage{booktabs}

\usepackage{graphicx}
\usepackage{xcolor}
\usepackage{textcomp}
\usepackage{txfonts}
\usepackage{amsmath}
\usepackage{subcaption}
\usepackage{xcolor} 
\usepackage{natbib,twoopt}
\usepackage[breaklinks=true]{hyperref}
\usepackage{url}
\bibpunct{(}{)}{;}{a}{}{,}             
\makeatletter
  \newcommandtwoopt{\citeads}[3][][]{\href{http://adsabs.harvard.edu/abs/#3}%
    {\def\hyper@linkstart##1##2{}%
     \let\hyper@linkend\@empty\citealp[#1][#2]{#3}}}
  \newcommandtwoopt{\citepads}[3][][]{\href{http://adsabs.harvard.edu/abs/#3}%
    {\def\hyper@linkstart##1##2{}%
     \let\hyper@linkend\@empty\citep[#1][#2]{#3}}}
  \newcommandtwoopt{\citetads}[3][][]{\href{http://adsabs.harvard.edu/abs/#3}%
    {\def\hyper@linkstart##1##2{}%
     \let\hyper@linkend\@empty\citet[#1][#2]{#3}}}
  \newcommandtwoopt{\citeyearads}[3][][]%
    {\href{http://adsabs.harvard.edu/abs/#3}
    {\def\hyper@linkstart##1##2{}%
     \let\hyper@linkend\@empty\citeyear[#1][#2]{#3}}}
\makeatother
\begin{document}

\title{POP-CORN: Validation of a new coronal hole detection tool based on neural networks}


    \author{K.H.P.  Henadhira Arachchige
           \inst{1}
           B. Perri\inst{1}
           \and
           A.S. Brun\inst{1}
          }

    \institute{Dept. d’Astrophysique/AIM, CEA/IRFU, CNRS/INSU, Université Paris et Paris-Saclay, 91191 Gif-sur-Yvette Cedex, France\\
              \email{kalpa.henadhiraarachchige@cea.fr}\inst{1}
             }



    \abstract
    {The properties and the spatial distribution of the large-scale structures of the solar corona determine the observed solar wind structure at 1~au. Coronal holes are a major source of fast solar wind, an important geo-effective component, and appear as large dark patches in extreme ultraviolet images. The solar observatories provide images of the solar corona at different wavelengths, enabling the identification of coronal hole morphology and other large-scale structures along a given line of sight. Mainly, the problem that arises is that there are models to do that, but few of them can work in real-time, separate coronal holes from other dark features, and are thought for comparison with models, and few of them are fully automatic.}
    {The main goal of this work is to develop an automatic threshold-based coronal hole detection tool across solar cycles 23, 24, and 25, using artificial intelligence. In this tool, the only user-input is the date, within solar cycles 23, 24, and 25, enabling the retrieval of the threshold value used to detect coronal hole contours in line-of-sight extreme ultraviolet images from SDO/AIA and SoHO/EIT.} 
    {We retrieve data that may affect the threshold value due to the change in contrast from the Heliophysics Events Knowledge database for the large-scale features such as active regions, solar flares, coronal mass ejections, and filaments, and then we engineer them, which allows us to train the neural network model (Prevision Of Phenomena through Coronal-hole Outline Recognition with Neural-network; POP-CORN). The model input comprises categorical features of large-scale structures in the solar corona, including their spatial distribution and additional properties, such as solar flare class by intensity. A neural network model (POP-CORN) was then trained to achieve higher accuracy. The model determines the threshold needed to detect coronal holes, allowing their boundaries to be identified automatically and consistently in extreme ultraviolet images from solar cycles 23, 24, and 25.}
    {To analyze the performances of our neural network model, we divided the predicted coronal hole results into different phases across the solar cycles 23, 24, and 25. Later, we compare them qualitatively and quantitatively with other state-of-the-art coronal hole detection tools.}
    {We conclude that the properties of large-scale structures affect the determination in coronal hole regions, and incorporating these properties manually into the training improves coronal hole detection. We find that POP-CORN performs well at detecting coronal hole contours, even when many bright features, such as active regions and solar flares, are present, which makes it hard for threshold-based methods to detect dark regions like coronal holes. In the future, we plan to integrate the coronal hole detection tool into a pipeline for validating solar wind models, creating a fully automated pipeline that provides a quantitative score for predictions.}

\authorrunning{K.H.P. Henadhira Arachchige}\titlerunning{Validation of a new coronal hole detection tool based on neural networks}

\maketitle
%

\section{Introduction}


Coronal Holes (CHs) are associated with open magnetic field lines that connect from the solar corona into the Parker-Spiral structure of the interplanetary magnetic field and extend throughout the heliosphere. These regions are the major source of the fast solar wind, the most geo-effective component of the solar wind. It is crucial to understand the properties of CH regions for effective space weather forecasting. These regions appear as dark patches in Extreme Ultra-Violet (EUV) and soft X-ray emissions \citet{Schwenn2006};\citet{Cranmer2009}, and the EUV images for solar corona can be retrieved from the solar observatories with different wavelengths, e.g., Atmospheric Imaging Assembly \cite[AIA,][]{AIA} on the Solar Dynamic Observatory (SDO), \cite{SDO},  SOlar and Heliospheric Observatory \cite[SOHO/EIT,][]{SOHO}, Solar Terrestrial Relations Observatory \cite[STEREO/SECCHI,][]{STEREO}, and Solar Orbiter \cite[SolO/EUI,][]{SolO}. CHs are often characterized by lower temperatures and densities than in the ambient corona. As the brightness of the solar corona mainly depends on the total energy input from photospheric magnetic activities \citep{Haruhisa_2025}, this is an indicator that CHs are regions of open magnetic field lines connected to photospheric regions of low magnetic activity.

Previous studies have demonstrated a wide range of methods for detecting CHs. The solar corona can be clearly observed in different wavelengths, which are 94$\AA$, 131$\AA$, 171$\AA$, 193$\AA$, 195$\AA$, 211$\AA$, 284$\AA$, 304$\AA$, 335$\AA$, in SDO/AIA and SoHO/EIT, which is also associated with different temperatures. Each EUV image from a different waveband reveals distinct CH morphologies, making qualitative identification of CH regions challenging. \citet{pyCATCH} discusses various approaches for identifying and extracting CH boundaries in SDO/AIA images.
These approaches are 1.) Single-wavelength, intensity-based threshold approach on EUV observations (\citet{Boucheron2016}, \citet{Hofmeister2017}, \citet{Heinemann2018}), 2.) Open field lines to characterize the CHs \cite[Potential Field Source Surface (PFSS),][]{PFSS}, \cite[Wang-Sheeley-Arge model (WSA model)][]{WSA}, \cite[MULTI-VP model][]{MULTI-VP}, 3.) Machine learning/neural network models e.g.,\citet{Illarionov2018}, \cite[pyCATCH][]{pyCATCH},\cite[CHRONNOS,][]{CHRONNOS}, 4.) CH extraction based on plasma properties \cite[Differential Emission Measure,][]{dem}, \citet{Hahn2011}. Another method is image segmentation \citep{Caplan2016}, which is more complex than intensity thresholding. In the intensity-thresholding technique, the image is converted into a binary image, highlighting the object of interest. However, in image segmentation, the method considers intensity, color, texture, shape, and other image-specific information.

Extracting CHs using the intensity-based threshold approach reveals that large-scale structures in the SC play a crucial role in identifying dark regions associated with CHs. The study from \citep{Hale1919}, \citep{McIntosh1990}, and \citep{Sammis2000} on Active Regions (ARs) confirmed that the intensity and the magnetic flux change according to the sunspot classification (Hale’s $\alpha$, $\beta$, and $\gamma$ classes). Similar studies reveal the dependencies of location and Solar Flare (SF) class on the intensity of the solar corona during erupting solar flares. The magnetic classification, $\delta$, introduced by \citep{Krunzel1960}, highlights the clear connection between SFs and magnetic connectivity \citep{warwick1966, mayfield1985, sammis2000, Moon2016}.
These studies confirm that most intense solar flares originate from the $\delta$ sunspot classification group. Also, the filament-like structures can be misinterpreted as CH regions in this method, potentially leading to false predictions \citep{Krista2009}, \citep{Reiss_2024}. 

Another effect on the SC intensity is coronal dimming, which can be observed in EUV and X-ray emissions that are associated with Coronal Mass Ejections (CMEs); however, these intensity changes only last for a short period, and it is challenging to distinguish between temperature and density changes in an EUV image during a dimming event. The relation between the dimming and the CMEs is not well understood \citep{SH_1997, harrison1997, zarro1999, HL_2000, H_2003}. Additionally, the survey by \citep{Chikunova_2023} investigated the spatio-temporal evolution of the dimming region.
Therefore, the NN models trained on EUV images (Convolutional Neural Network: CNN) may not adequately capture the regions and properties of large-scale structures (e.g., solar flares, filaments, and darker quiet-sun regions) in the SC and may require manual incorporation into training. When considering the various phases of the solar cycle, the CH morphology changes are clearly observable, and the structure is visible across the rising, declining, maximum, and minimum phases. Therefore, the intensities of the EUV images play a vital role in identifying CH boundaries when training NN models within the solar cycle.

In the article by \citet{Reiss_2024}, a comprehensive discussion of various CH detection schemes is provided. However, CH detection may be challenging due to effects caused by 1). Stray light from nearby regions, 2.) Instrument effects, and 3.) Limb brightening (\citet{Verbeeck_2014}; \citet{Caplan2016}). Here, in this paper, we introduce a different approach in training the NN model, by manually implementing the properties and the locations of the large-scale structures of the SC into a NN model, in order to improve further CH identification. Therefore, the inputs are categorical or binary (0/1), and the output is a linear threshold used to detect CH segmentations in LOS-EUV images. 

In this paper, we present an automated tool (Prevision of phenomena through coronal hole outline recognition with neural network: POP-CORN) for detecting CH boundaries in a LOS-EUV image (SDO/AIA or SoHO/EIT) of the SC over solar cycles 23, 24, and 25 for a user-defined time period. We describe our modelling approach and setup in Section~\ref{sec:model} and provide the coronal hole contour predictions with qualitative and quantitative comparisons in Section~\ref{sec:results}. We then discuss how it compares to other CH detection techniques in Section~\ref{sec:discussion}, and conclude our findings in Section~\ref{sec:Conclusions}.

\section{AI Model}\label{sec:model}

In this section, we will describe the AI model and the process for to identify CH boundaries. This model is based on the properties and spatial distribution of the large-scale structures of the SC. We use data from the Heliophysics Event Knowledge (HEK) database, along with extreme ultraviolet (EUV) data from the MEDOC
and FIDO database. In the following subsections, we describe how we identify the CH boundaries in Section~\ref{subsec:Identifying_the_CH_boundaries}, the relation between the threshold values in SDO \& SOHO in Section~\ref{subsec:Relation_between_the_threshold_values}, the data preprocessing and engineering in Section~\ref{subsec:feature_eng}, model training in Section~\ref{subsec:NN_model}, and evaluation processes used to assess its performance in Section~\ref{subsec:Model_Validation}.

\begin{figure}[htpb]
\centering
\begin{subfigure}{0.4\linewidth}
    \centering
    \includegraphics[width=\linewidth]{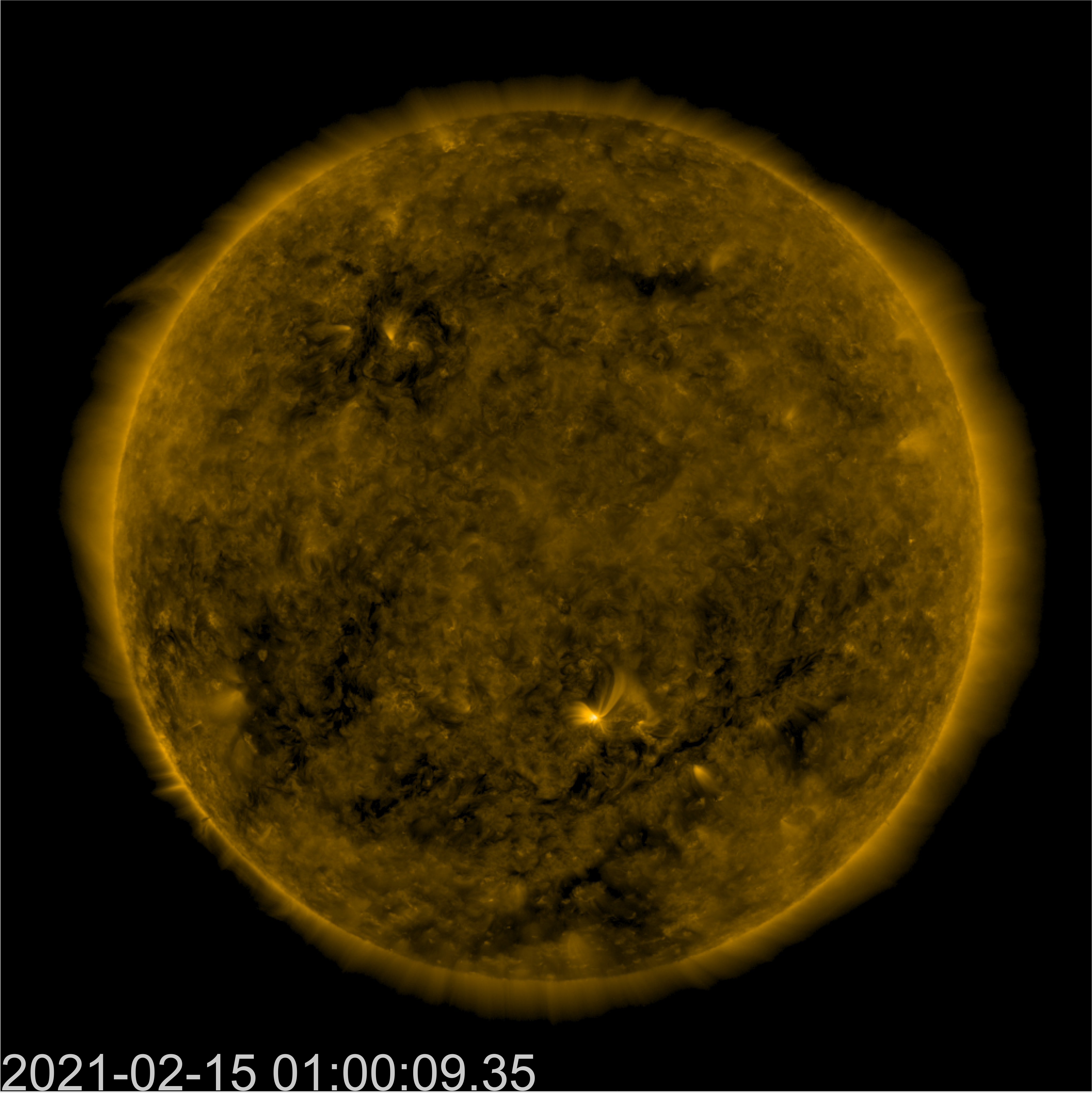}
    \caption{SDO/AIA \textit{171{\AA}}}
\end{subfigure}
\begin{subfigure}{0.4\linewidth}
    \centering
    \includegraphics[width=\linewidth]{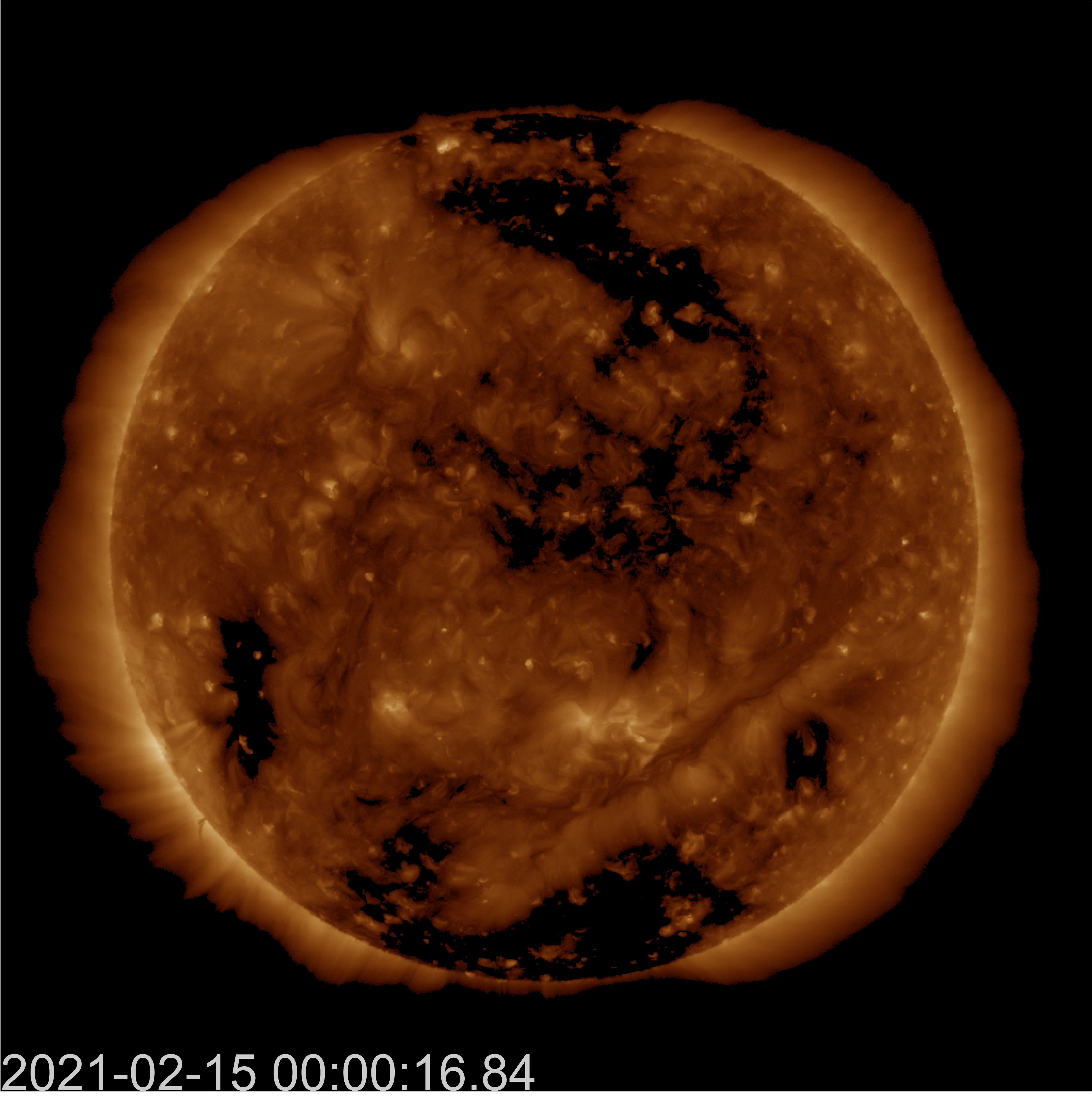}
    \caption{SDO/AIA \textit{193{\AA}}}
\end{subfigure}


\begin{subfigure}{0.4\linewidth}
    \centering
    \includegraphics[width=\linewidth]{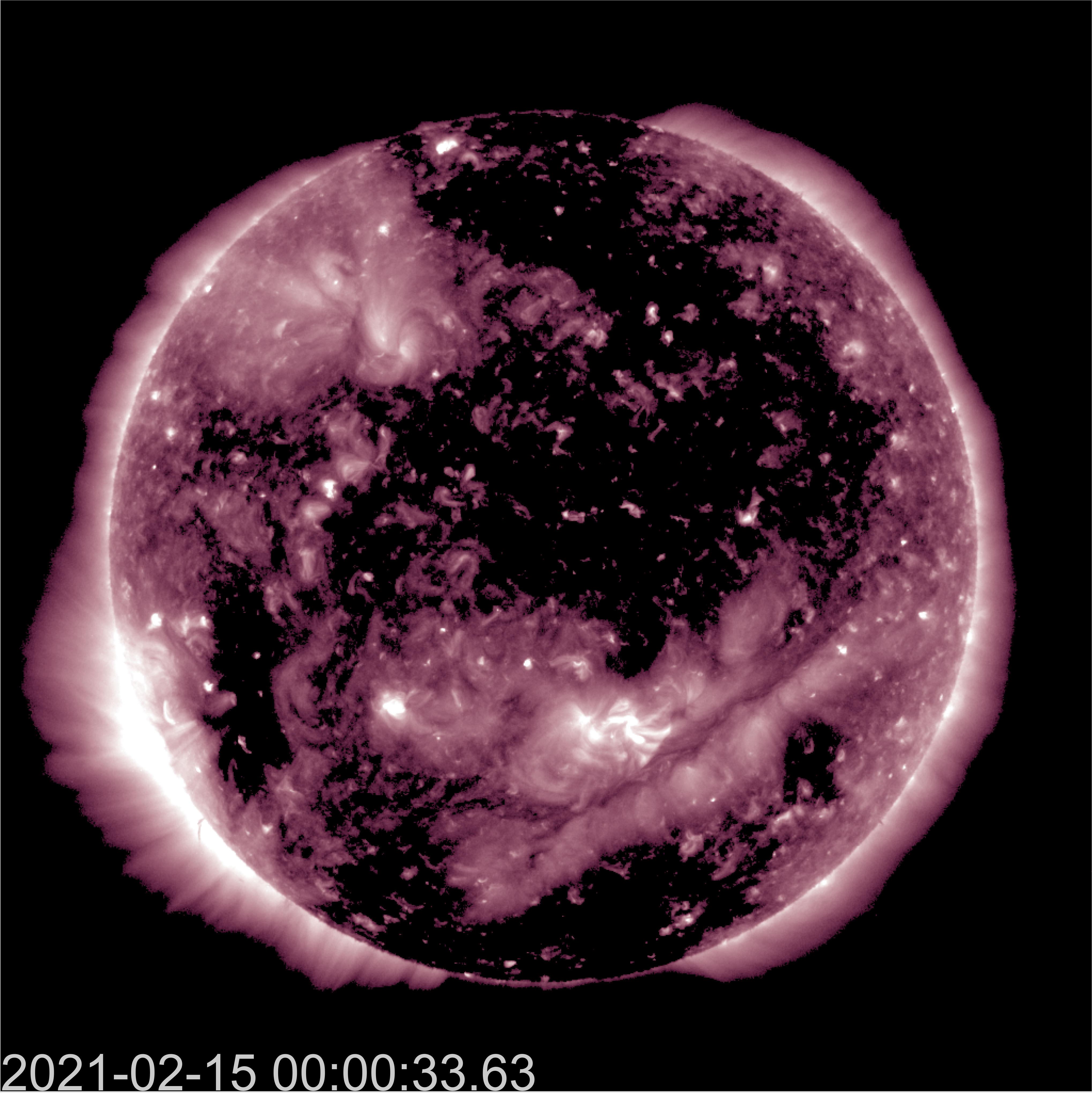}
    \caption{SDO/AIA \textit{211{\AA}}}
\end{subfigure}
\begin{subfigure}{0.4\linewidth}
    \centering
    \includegraphics[width=\linewidth]{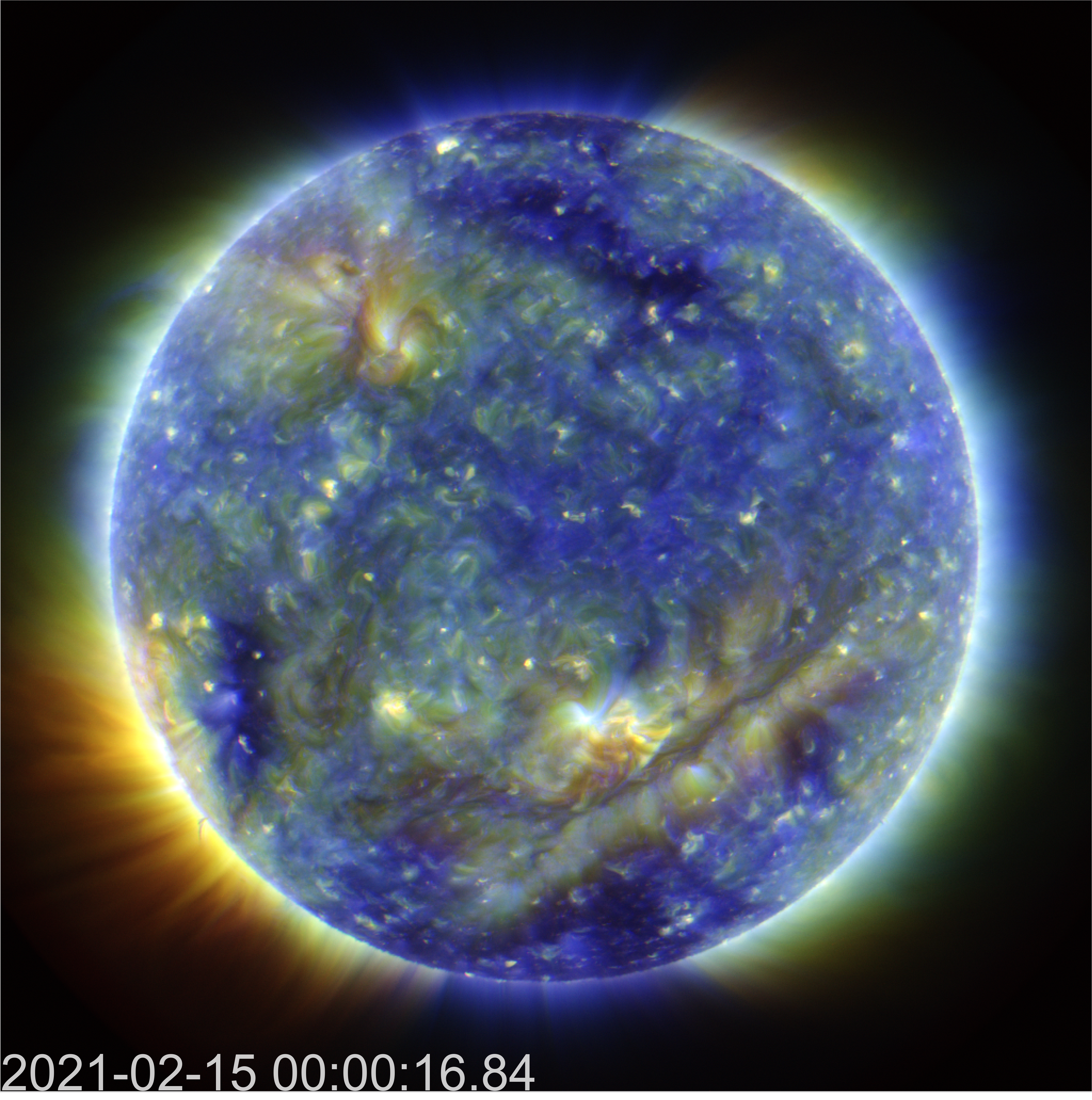}
    \caption{SDO/AIA Composite}
\end{subfigure}

\caption{EUV images of the Sun from SDO/AIA for the time period of 2021:02:15 for the wavelengths \textit{171{\AA}}, \textit{193{\AA}}, \textit{211{\AA}}, and for the composite image.}
\label{fig:CR2240_wavelength_comp}
\end{figure}

\subsection{Identifying the Coronal Hole boundaries}
\label{subsec:Identifying_the_CH_boundaries}

To identify the CH boundaries, we first select LOS-EUV images from solar cycles 23, 24, and 25. Figure~\ref{fig:CR2240_wavelength_comp} illustrates the coronal hole morphologies observed from different wavelengths (171$\AA$, 193$\AA$, 211$\AA$), and the composite of 171{$\AA$}+193{$\AA$}+211{$\AA$}, for the time period of \textit{2021:02:15}. Here, we clearly observe the difference in the coronal hole morphologies where; 171$\AA$, and 211$\AA$, EUV images do not provide clear information on dark regions such that one cannot easily distinguish CHs from other features, whereas by contrast both the 193$\AA$, and the composite image provide the details of the coronal holes better at the probed coronal height, even though they are taken at the same time. From a qualitative perspective, we find that the best option for identifying the CHs is the composite image at wavelengths \textit{171{$\AA$}} + \textit{193{$\AA$}} + \textit{211{$\AA$}}. 

To identify the CH boundaries, we selected a single LOS-EUV image from SDO/AIA at 193$\AA$ and a composite image at the same time, representing a Carrington Rotation (CR) period, for training. We observe dark regions that match the exact CH regions seen with the naked eye. Then we identify the optimal threshold value to determine the CH regions of the selected LOS-EUV for each CR period within the above cycles, this process is discussed later in this section. We use the \textit{opencv-python} and \textit{sunpy} libraries to trace the boundaries of the dark regions and mask the solar disk, respectively. We want to be able to trace the CH profiles for both SoHO/EIT \textit{195{$\AA$}} (representing the minimum and rising phase of solar cycle 23) and SDO/AIA \textit{193{$\AA$}} (representing solar cycles 24 and 25). However, to observe the CH regions in a specific LOS-EUV image, we need to distinguish them from other dark features visible to the naked eye. For example, structures such as \textit{filaments} may be misinterpreted as CHs. Therefore, we not only take into account the contrast of the structures, but also their shape in our eye identification, as filaments tend to be more elongated than coronal holes.
We then binarize the pixel data to obtain a gray-scale image and apply a threshold to identify dark regions. We identified the optimal threshold to match the CHs using both qualitative and quantitative metrics (as described later; see Appendix~\ref{sec:HT} for more details) and subsequently collected thresholds for SDO/AIA and SOHO/EIT EUV images.

For the quantitative method, we also compare our CH predictions with the area in the HEK database. Ultimately, we determined the optimal threshold for each LOS-EUV image from SDO/AIA and SOHO/EIT across solar cycles 23, 24, and 25. 

\begin{figure}[htpb]
\centering
\begin{subfigure}{0.45\linewidth}
    \centering
    \includegraphics[width=\linewidth]{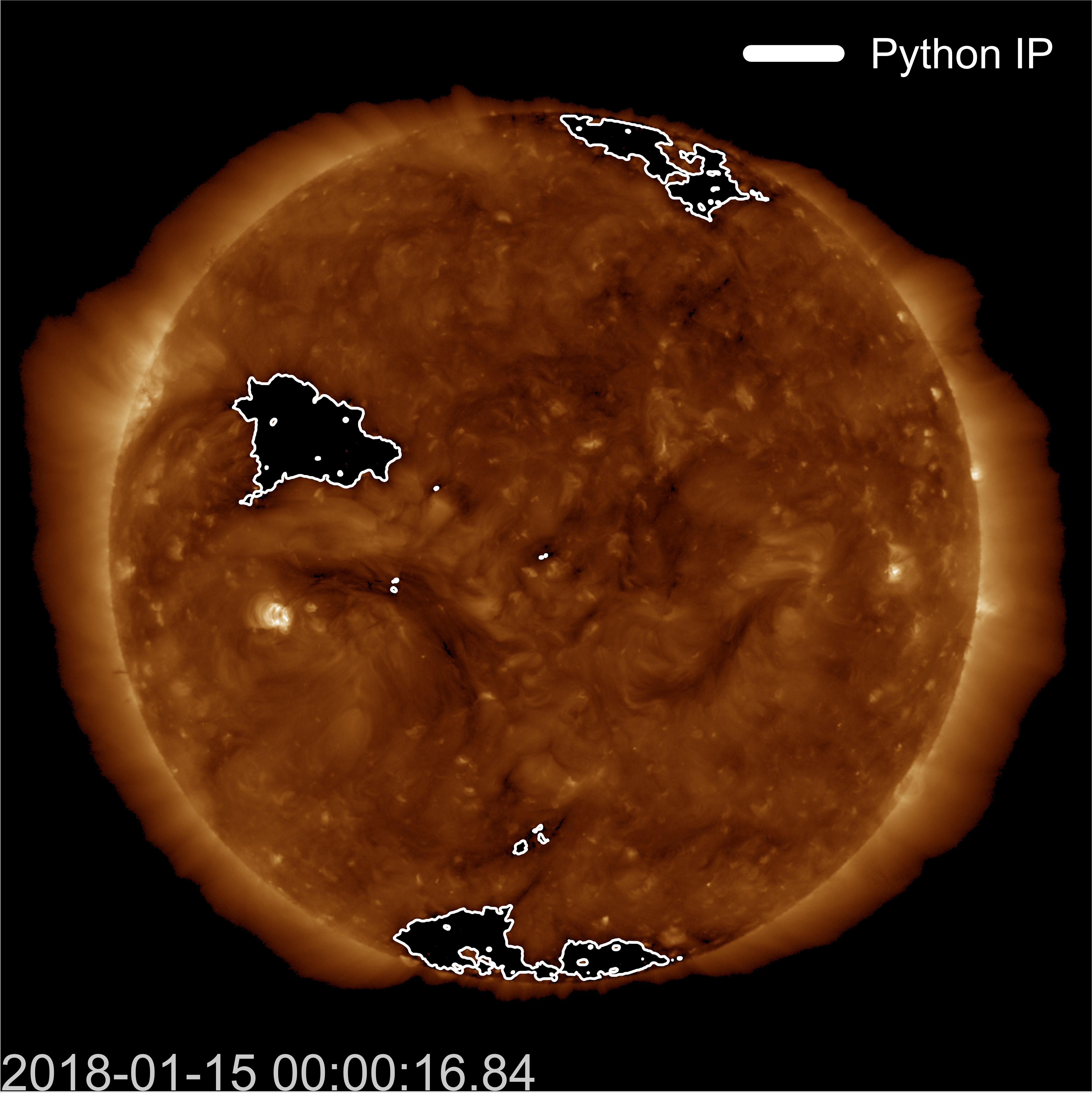}
\end{subfigure}
\begin{subfigure}{0.454\linewidth}
    \centering
    \includegraphics[width=\linewidth]{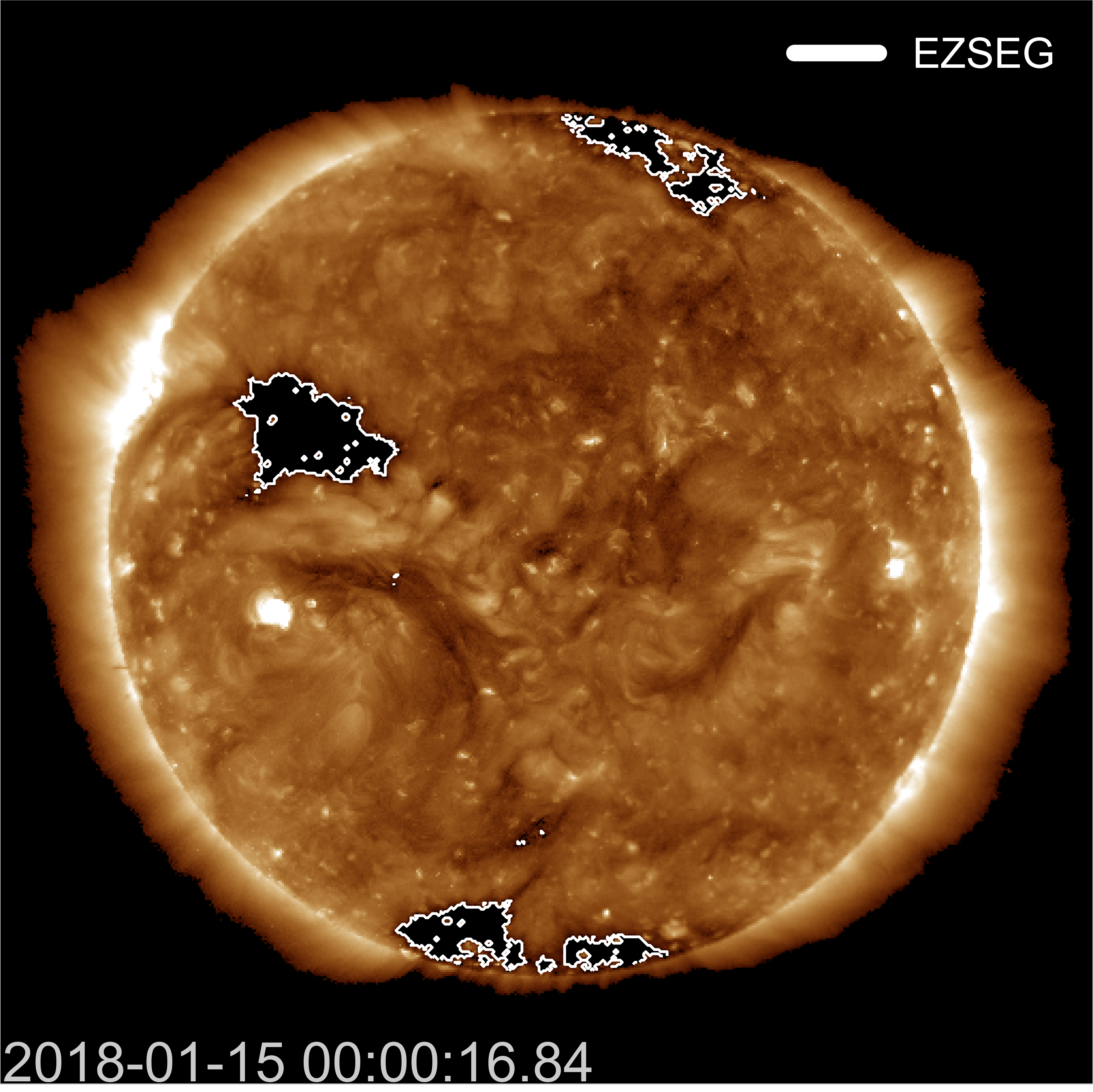}
\end{subfigure}
    \caption{Comparison for threshold value from our image processing technique (left) and from the EZSEG algorithm (right) for the date \textit{2018-01-15 00:00:16.84} for SDO/AIA \textit{193{$\AA$}}.}
    \label{fig:EZSEG_comparison}
\end{figure}

Figure~\ref{fig:EZSEG_comparison} illustrates the CH detection with a threshold value, the left image is from the Python image processing, and the right is from the \textit{EZSEG} algorithm. We clearly see that the Python image processing captures the CHs near the poles, as does the \textit{EZSEG} algorithm, and it doesn't mislead in detecting other dark regions, such as filaments.

\subsection{Relation between the threshold values in SDO \& SOHO}
\label{subsec:Relation_between_the_threshold_values}

\begin{figure}[htbp]
    \centering
    \includegraphics[width=0.48\textwidth]{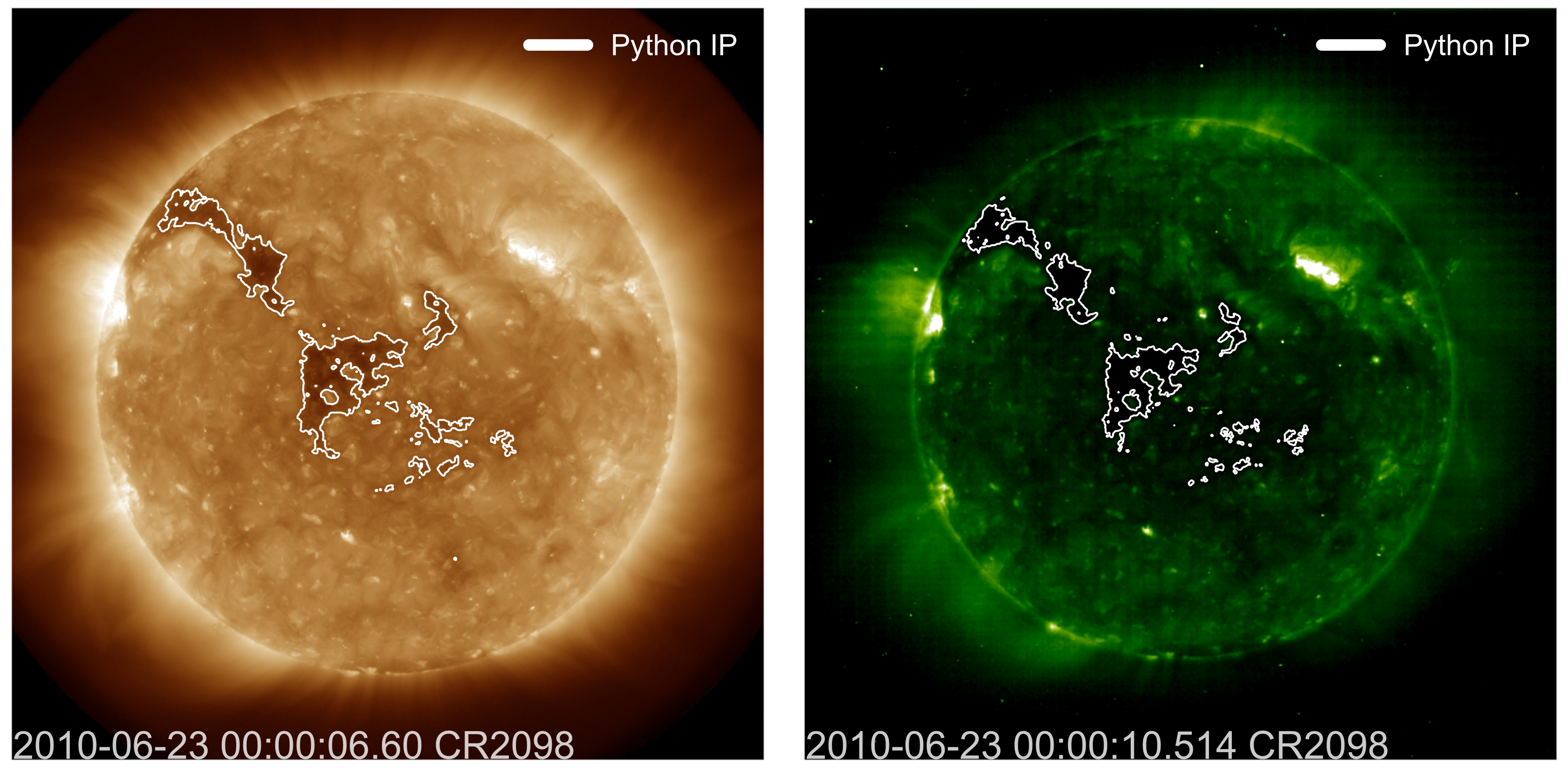} 
    \caption{Illustration of the relationship between SDO/AIA 193$\AA$ (left) and SOHO/EIT $195\AA$ (right) threshold values for \textit{2010-06-23 00:00:06.60} \textbf{1024 X 1024} resolution.}
    \label{fig:SDO_SOHO}
\end{figure}

The model is developed to predict daily CH forecasts and retrieve CH boundaries during solar cycles 23, 24, and 25. Since the SDO/AIA observatory does not provide images before the CR2098 period \textit{(2010-06-15 16:29:54)}, we use SOHO/EIT \textit{195{$\AA$}} waveband images to train the model and ensure it is complete for solar cycles 23, 24, and 25. However, the image intensities differ across observations, and in the SDO/AIA images, the pixel scale is 0.6 arcsec/pixel, and it provides fine detail and resolves thin coronal loops, small-scale brightening, and fine filament structures, but in SOHO/EIT images, the pixel scale is 2.6 or 5.26 arcsec/pixel, with structures and small features being blended or unresolved. Therefore, it is difficult to use the same threshold values to identify CH boundaries as in the SDO/AIA images, due to their shorter wavelengths and greater sensitivity to small threshold variations. We train a simple additional linear NN model to learn the relationship between the threshold values for CH boundary detection in the SDO/AIA and SOHO/EIT EUV images. In Figure \ref{fig:SDO_SOHO}, we see an example of the relation for a specific time period; the model is implemented in the main model to combine and produce predictions for CHs across all three solar cycles.

We also perform a quantitative comparison to validate the threshold, using pixel coordinates within the contours. Refer to Appendix~\ref{sec:appendix_SDO-SoHO} for the quantitative comparison.

\subsection{Feature engineering}
\label{subsec:feature_eng}

In this subsection, we discuss how we handle the data for training the main NN model. We retrieved data for large-scale structures of the SC, which are, Active Regions (AR), Solar Flares (SF), Coronal Mass Ejections (CMEs), Filaments (FI), Filament Eruption (FE), Coronal Dimming (CD), Emerging Flux (EF), and Phase of the Solar Cycle from the \textit{HEK} database and the sunspot number (SSN) from the \textit{SILSO} observatory. These structures affect the contrast of the EUV image, thus the threshold value to determine the CH boundaries. We organize large-scale features by spatial distribution and their properties. For spatial distribution, we categorized the events that occurred based on dividing the LOS-EUV image solar disk surface into quadrants (please refer to Appendix~\ref{sec:model_input_features}) to view how we categorized the large-scale structures based on the spatial distribution.) This enhances the model training by capturing the spatial distribution and using the intensity of the EUV image to determine CH regions. For SF, we categorized them by class based on their intensity from GOES data (e.g., X, M, C, A, B), (please refer to Appendix~\ref{subsec:GOES}) and the active regions from which the sunspot classification group emerged from the Mount Wilson magnetic classification (please refer to Appendix~\ref{subsec:MountWilson}). As in the above cases, we categorized the other three features (FI, FE, CD, and EF) by the quadrant of the solar disk where each event occurred and by the phase of the solar cycle. In total, we have 93 distinct categorical features for training the model. 

\subsection{Neural Network (NN) Model (POP-CORN)}
\label{subsec:NN_model}

Here, we trained a NN model with hidden layers. The architecture of the NN configuration is illustrated in Figure~\ref{fig:NN_Config}, and the activation function for each layer is illustrated in the diagram. As mentioned above in Section~\ref{subsec:feature_eng}, we used all 93 categorical features as inputs to our NN model, and the output is a single threshold value. We use the \textit{sigmoid} function as the activation function for the input layer, and the \textit{linear} function for the output layer. For the hidden layers, we use the \textit{ReLU} activation function. To improve model convergence, we add batch normalization at each layer and a dropout layer with a 0.3 dropout rate, which helps prevent overfitting to the training data. 

The number of hidden layers is adjusted to minimize the loss for both the training and validation datasets. We prioritized avoiding both overfitting and underfitting. The loss is calculated from the \textit{Mean Squared Error (MSE)}. We use Mean Absolute Error \textit{(MAE)} as the metric, and \textit{Adaptive Moment Estimation (Adam)} as the optimizer in the NN model to minimize the error in the loss function and maximize the efficiency.

\subsection{Model Validation}
\label{subsec:Model_Validation}

\begin{figure}[htbp]
    \centering
    \includegraphics[width=0.27\textwidth]{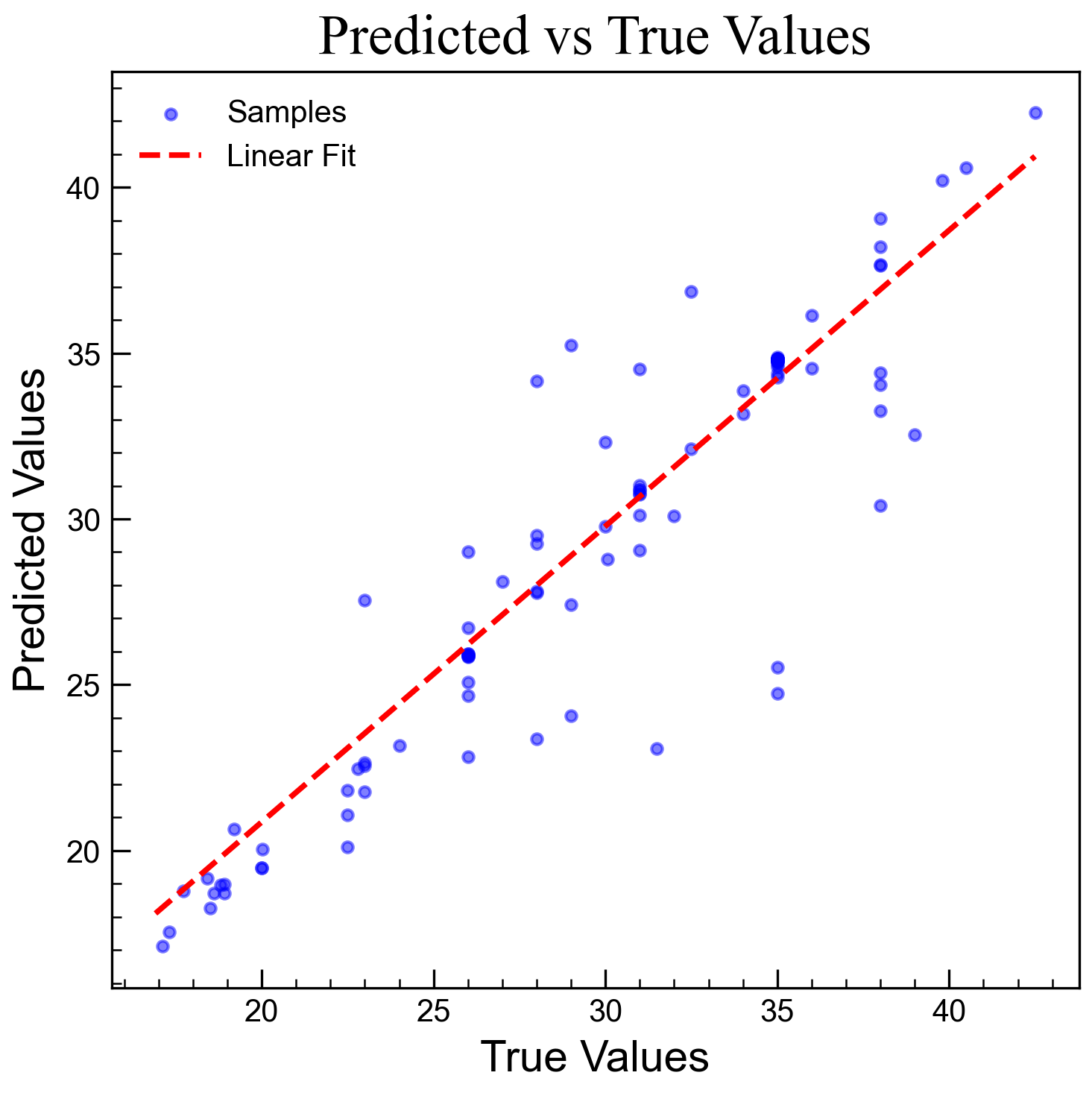} 
    \caption{Figure showing the correlation between predicted data and true values. The red line denotes the fitted linear regression model.}
    \label{fig:corr}
\end{figure}

\begin{figure}[htbp]
    \centering
    \includegraphics[width=0.30\textwidth]{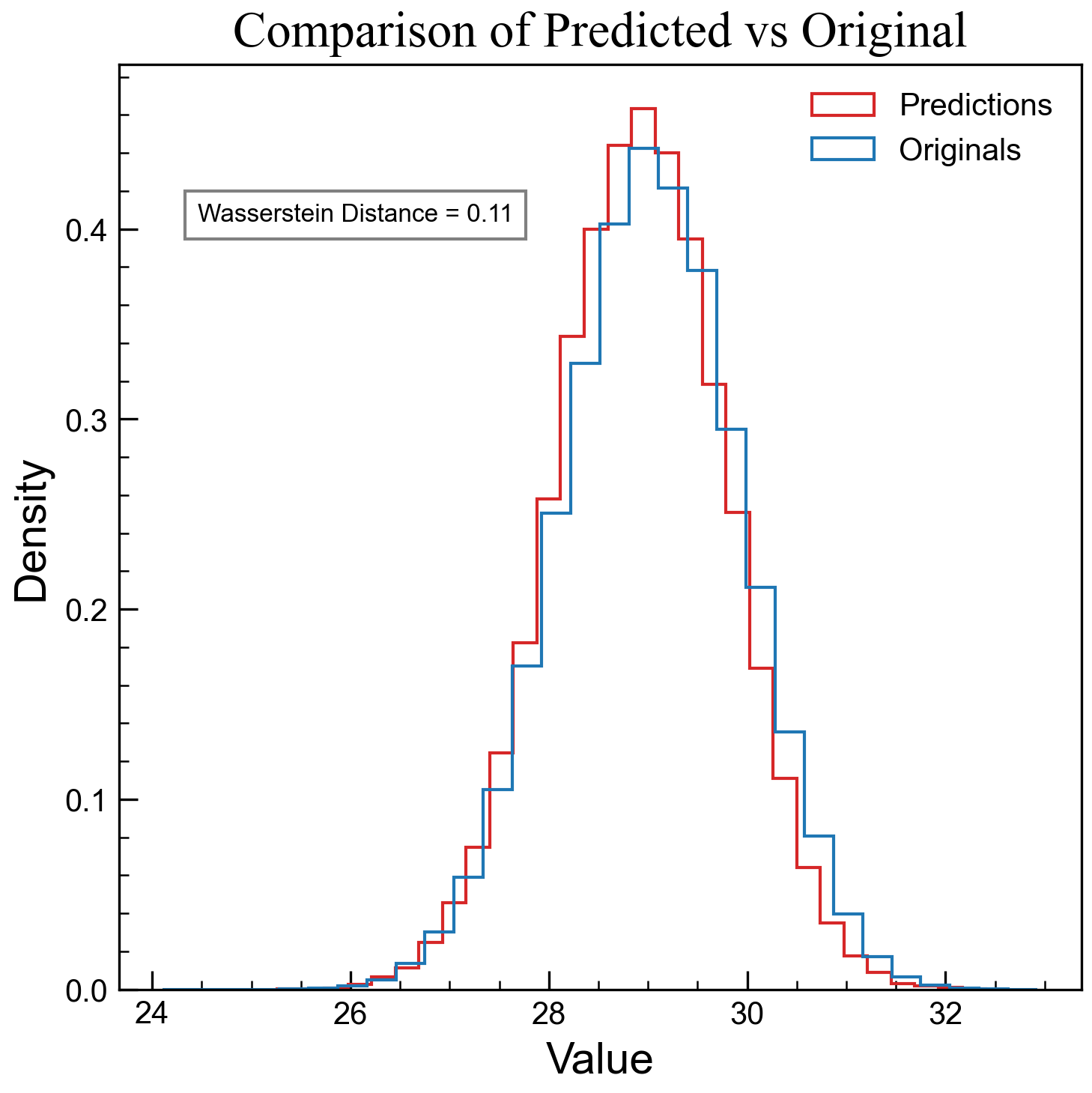} 
    \caption{Figure showing overlaid histograms of predicted and true values. The Wasserstein distance quantifies the difference between the two distributions.}
    \label{fig:hist}
\end{figure}

In this section, we discuss the validation of the POP-CORN model. Here, we use 75\% of the data from the period June 2010 to July 2022 (for the model of SDO/AIA EUV images), which we processed in Section~\ref{subsec:feature_eng}, to train the model, and the remaining 25\% to validate it. Figure~\ref {fig:corr} illustrates the correlation between the predicted threshold values from POP-CORN and the validation data; we observe a Pearson correlation of 0.94, which is very good, as the value is very close to 1.  
To confirm the validation, as illustrated in Figure~\ref{fig:hist}, we perform a histogram comparison for the validation data against the POP-CORN predicted threshold values. We also conducted the Wasserstein distance\footnote {\href{https://docs.scipy.org/doc/scipy/reference/generated/scipy.stats.wasserstein_distance.html}{SciPy documentation for Wasserstein distance}} statistical test to validate the result, here we found that the statistical test metric value is 0.11, which implies the difference between the distribution of predicted and the true values is small. For statistical comparisons using the Wasserstein distance, please refer to \citet{HenadhiraArachchige2022}.

\begin{figure}[htbp]
    \centering
    \includegraphics[width=0.32\textwidth]{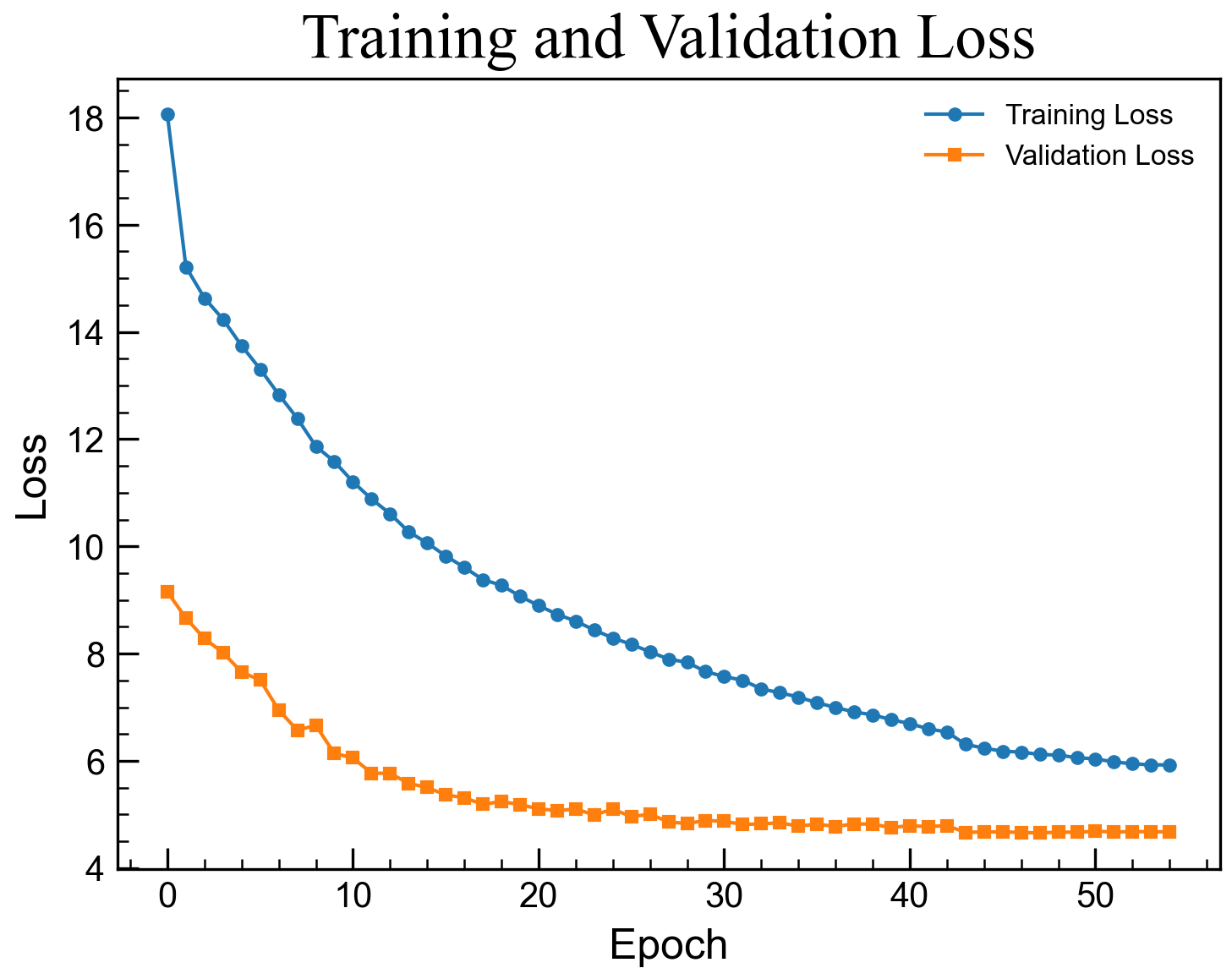} 
    \caption{Plot of the training (blue) and validation (orange) loss curves, illustrating model performance over epochs.}
    \label{fig:loss}
\end{figure}

Figure~\ref{fig:loss} shows the loss curves for training and validation during the model training. Here, the main idea is to present the model performance based on the \textit{Mean Squared Error (MSE)}. We observe that at the beginning of the training, the MSE (or loss) is high, and as training converges, the MSEs for training and validation converge, which means that the model generalizes to both training and validation data (preventing overfitting and underfitting).

Here, an epoch denotes one pass over the entire dataset through the neural network during training. (50 epochs: NN model sees all the training datasets 50 times.)

\section{Results}
\label{sec:results}

In the results section, we present the CH estimations for each phase of the solar cycles 25, 24, and 23. Here, we qualitatively examine the CHs produced at each cycle and the statistical performance of the model. Please refer to the Appendix~\ref{sec:HT} for the equations and the hypotheses of the Hotelling's $T^2$ test. We also compare the CH predictions from POP-CORN with the ground truth (GT), which is the CHs observed at an optimal threshold in the EUV image by eye. From the results of Hotelling's $T^2$ test \textbf{fail to reject $H_0$}, it implies that the model agrees well statistically with the GT.

\subsection{Solar Cycle 25}
\label{subsec:solar_cycle_25}

\begin{figure}[htpb!]
\centering
\begin{subfigure}[b]{0.46\linewidth}
    \centering
    \includegraphics[width=\linewidth]{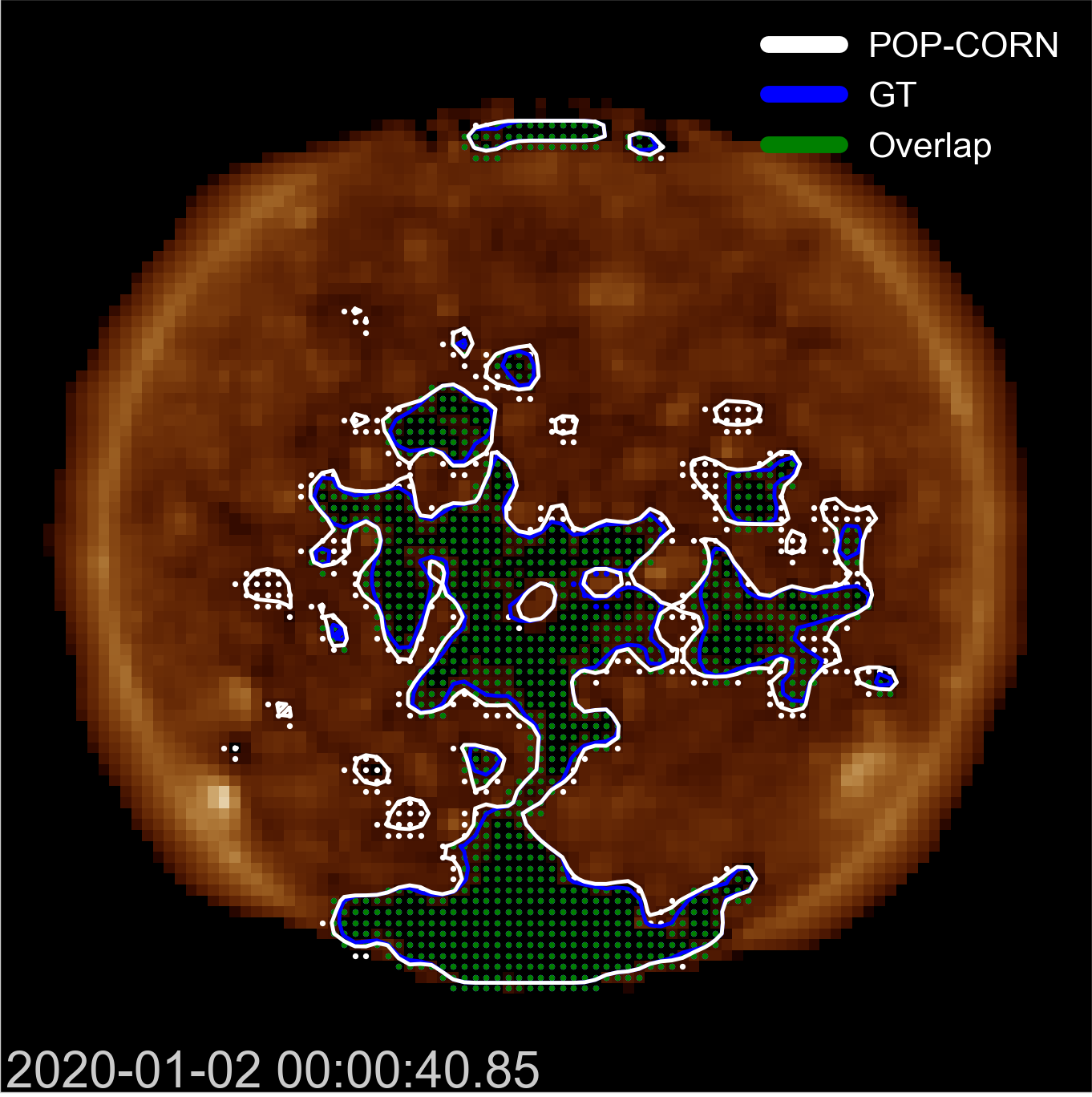}
    \caption{CR2225 (Minimum)}
    \label{fig:SC25-min-CR2225}
\end{subfigure}
\begin{subfigure}[b]{0.46\linewidth}
    \centering
    \includegraphics[width=\linewidth]{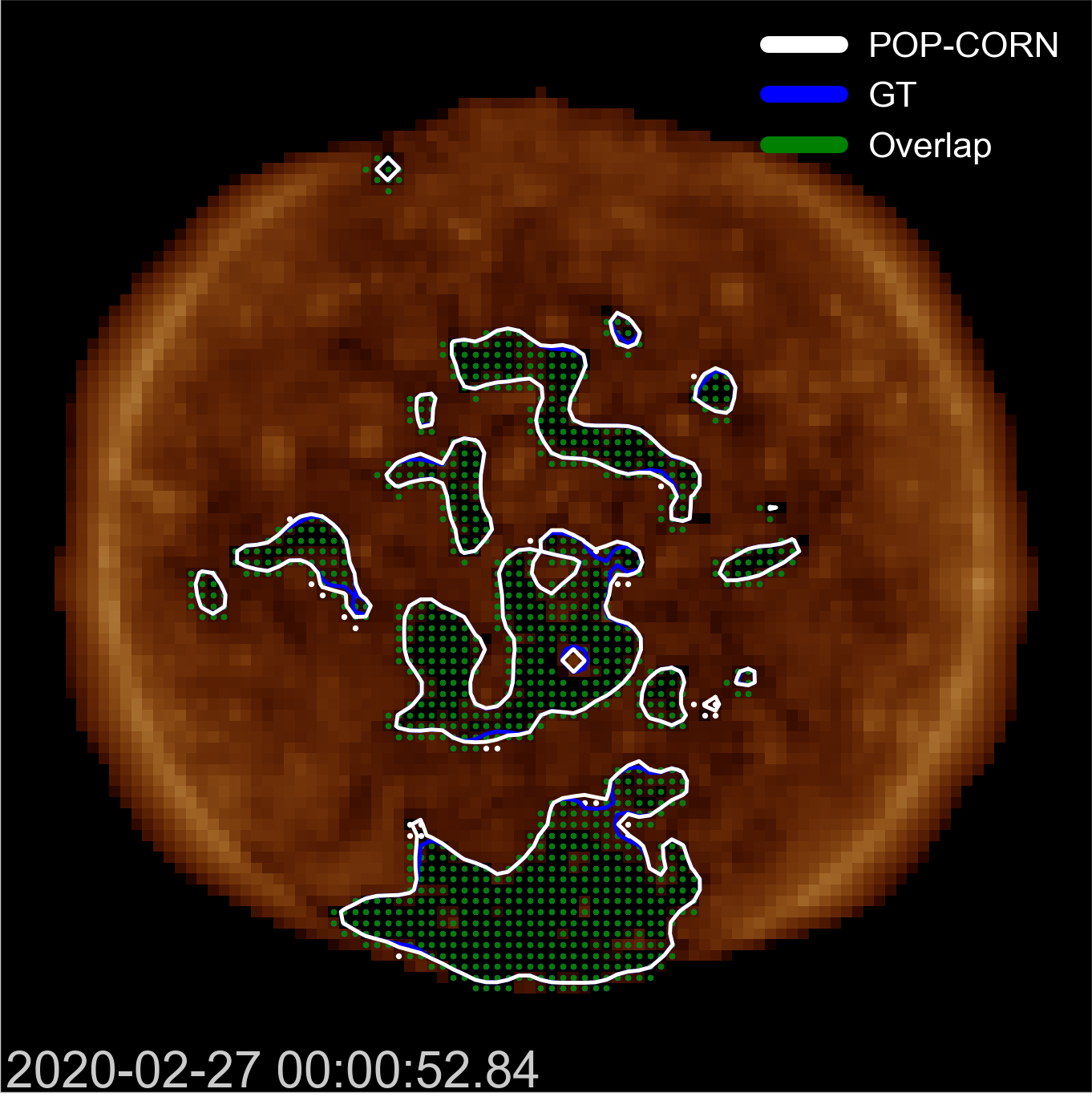}
    \caption{CR2227 (Minimum)}
    \label{fig:SC25-min-CR2227}
\end{subfigure}
\begin{subfigure}[b]{0.46\linewidth}
    \centering
    \includegraphics[width=\linewidth]{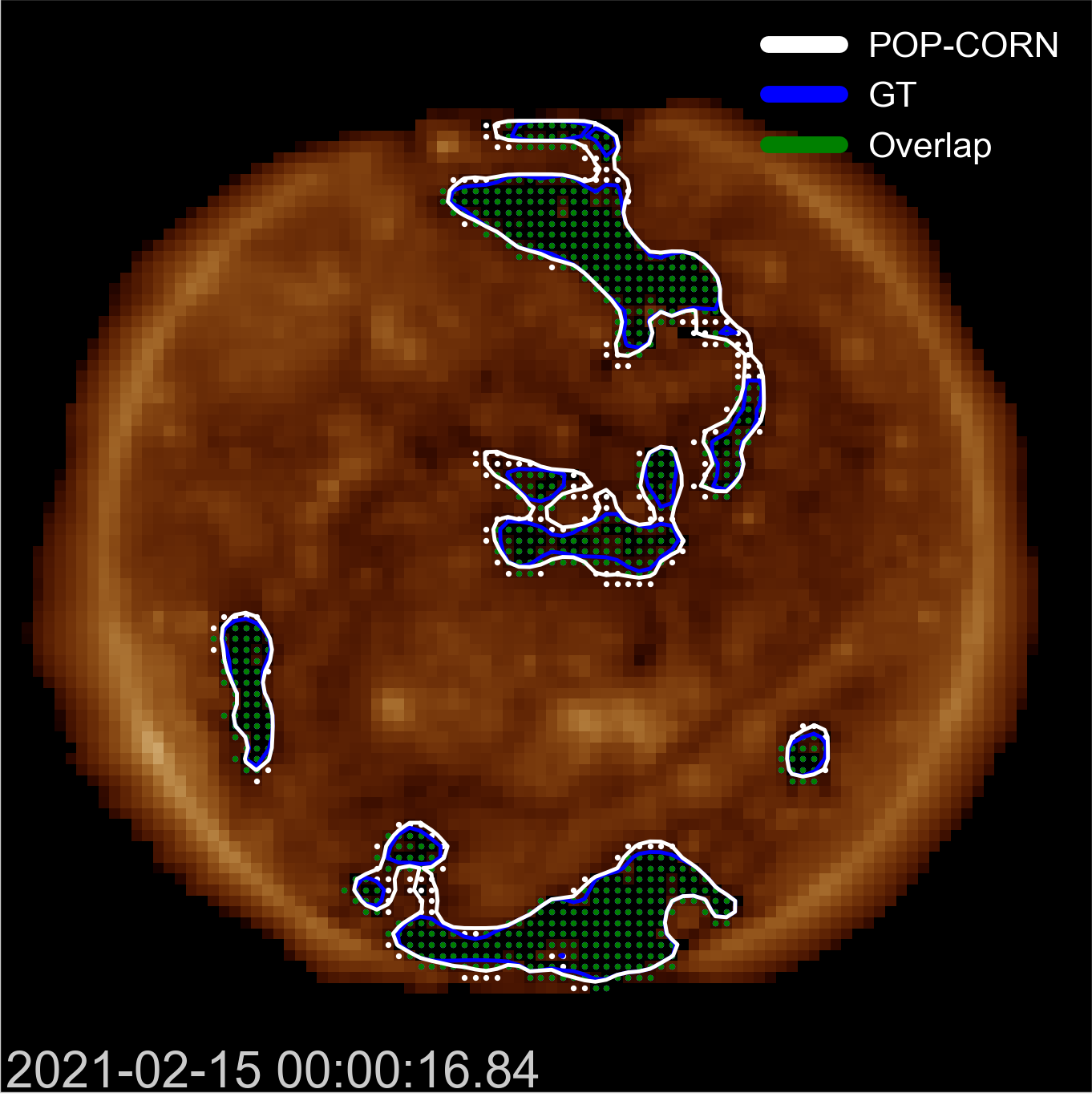}
    \caption{CR2240 (Rising)}
    \label{fig:SC25-ris-CR2240}
\end{subfigure}
\begin{subfigure}[b]{0.46\linewidth}
    \centering
    \includegraphics[width=\linewidth]{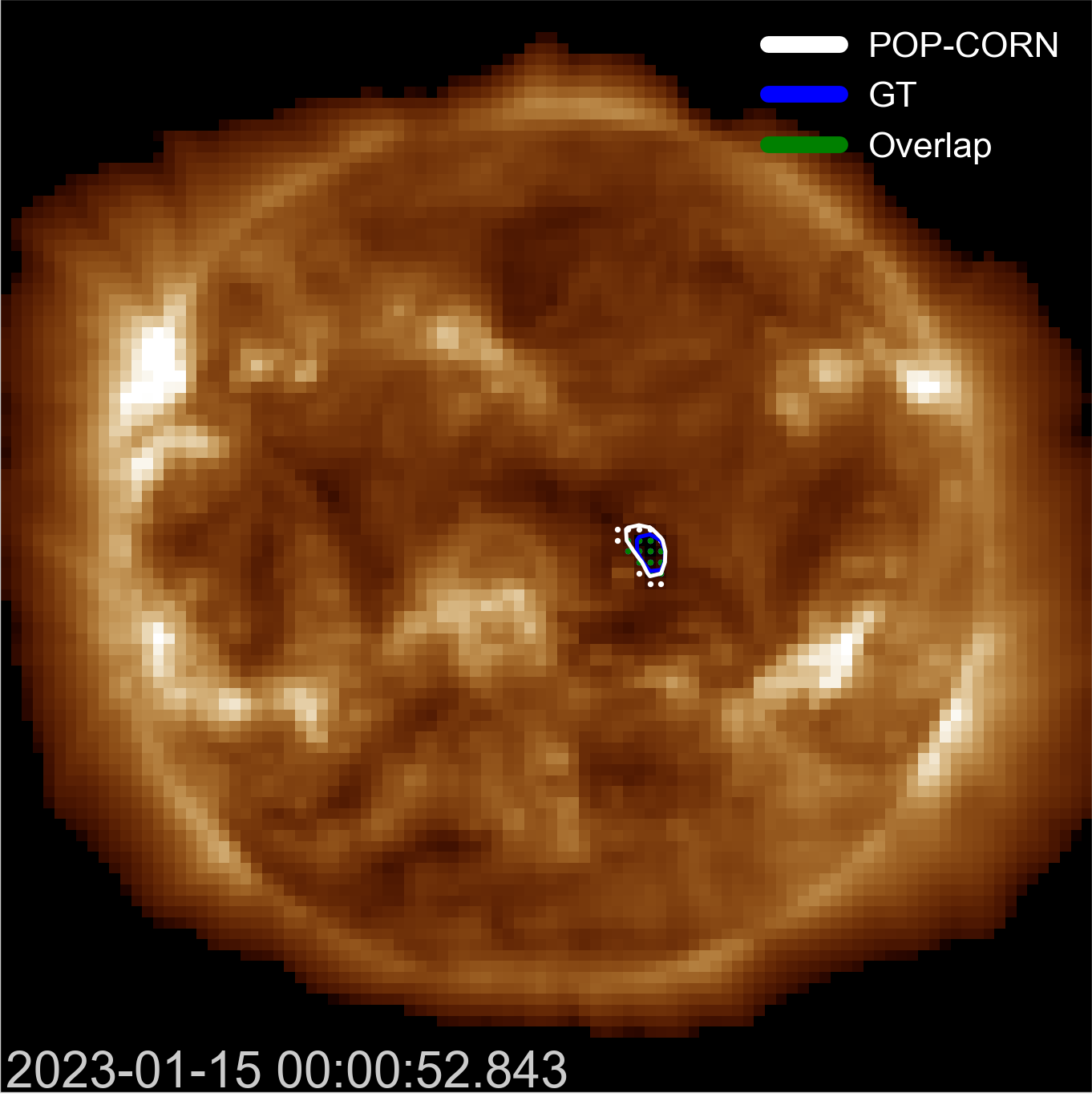}
    \caption{CR2266 (Rising)}
    \label{fig:SC25-ris-CR2266}
\end{subfigure}
\begin{subfigure}[b]{0.46\linewidth}
    \centering
    \includegraphics[width=\linewidth]{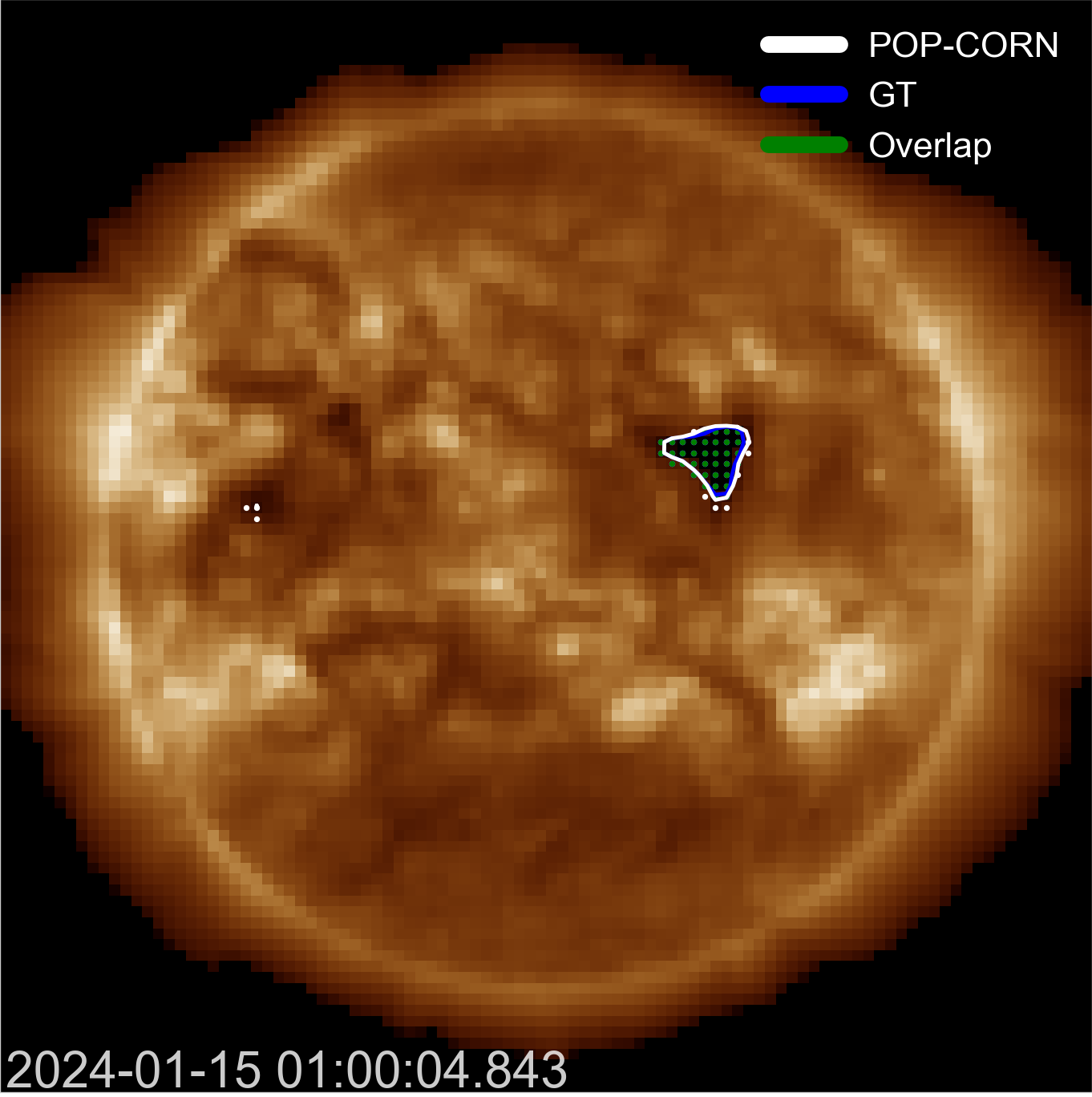}
    \caption{CR2279 (Maximum)}
    \label{fig:SC25-max-CR2279}
\end{subfigure}
\begin{subfigure}[b]{0.46\linewidth}
    \centering
    \includegraphics[width=\linewidth]{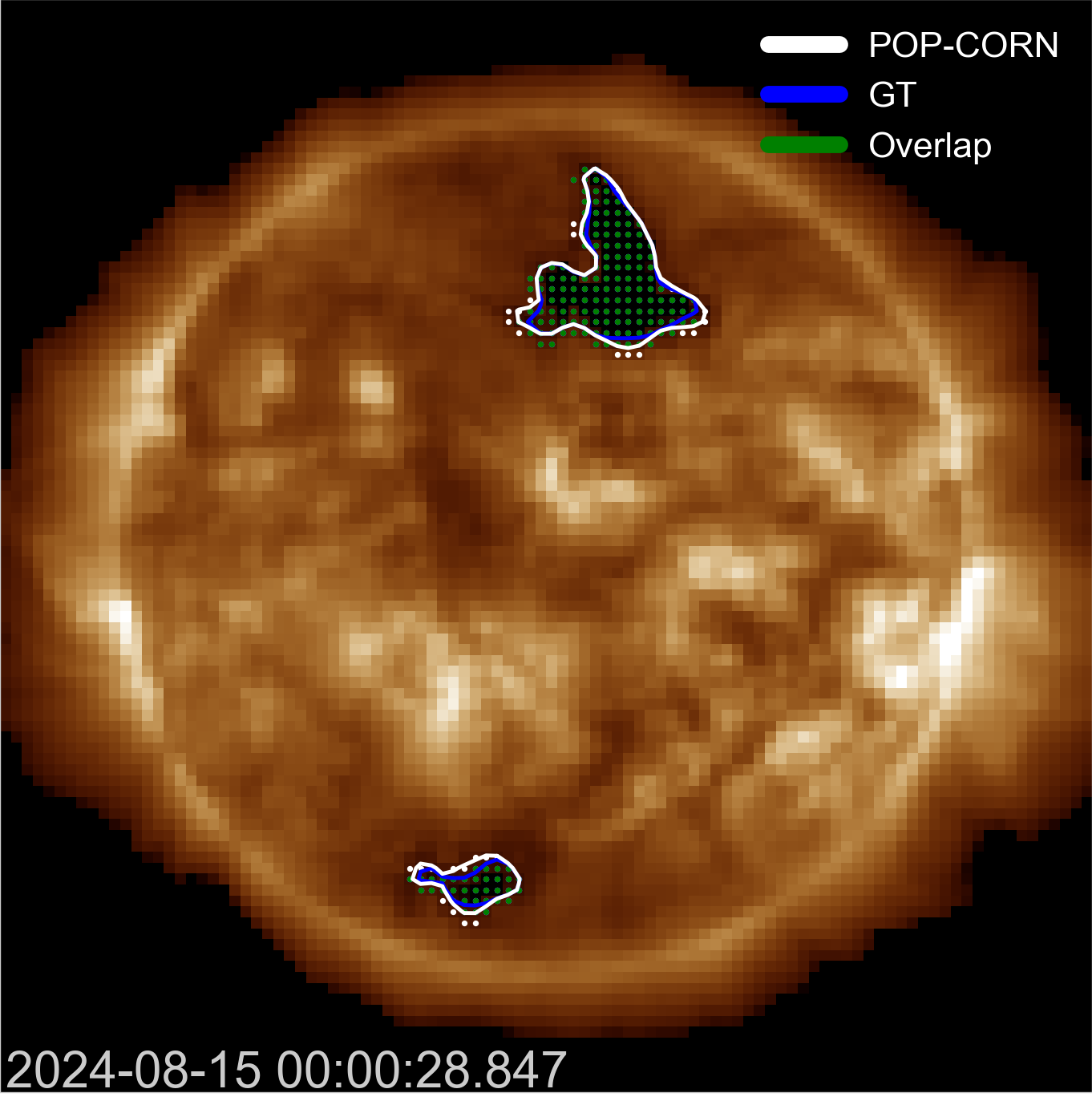}
    \caption{CR2287 (Maximum)}
    \label{fig:SC25-max-CR2287}
\end{subfigure}
\caption{Predicted coronal hole (CH) segmentation across different phases of Solar Cycle 25 (minimum, rising, and maximum) using EUV images from SDO/AIA at 193$\AA$.}
\label{fig:SC25-all-phases}
\end{figure}

The following images (Figure~\ref{fig:SC25-all-phases}) represent the CH predictions from POP-CORN for the solar minimum, rising, and solar maximum, compared with the pixel coordinates within the contours from the GT. The POP-CORN estimated contours are shown in white, while the GT contours are shown in blue.  The pixels within the estimated coronal holes are colored green when they match the GT, and remain white if they are false positives. Pixels are colored in blue when they are false negatives (GT not reproduced by POP-CORN). These results demonstrate the model’s ability to detect CH boundaries across different phases of the solar cycle, both during periods of high solar activity and at solar minimum. Here, we present the results for Solar Cycle 25, for the solar minimum, rising, and solar maximum phases. From a qualitative perspective, we observe that the model performs well in predicting CHs, and we further conduct a statistical comparison between the predicted CHs and the GT (Table~\ref{tab:sc25_Hotelling}).

In Figures~\ref{fig:SC25-min-CR2225}, \& \ref{fig:SC25-min-CR2227}, we present the CH estimations for the minimum phase of Solar Cycle 25. At a minimum of activity, we have a low contrast of the pictures, due to the absence of active regions. This also leads to large and fragmented coronal holes. We can see that we have an almost perfect match between the GT and POP-CORN contours. Additionally, based on the statistical test presented in Table~\ref{tab:sc25_Hotelling} (lines 1 and 2), the CH estimations shown in Figures~\ref{fig:SC25-min-CR2225} (CR2225) and \ref{fig:SC25-min-CR2227} (CR2227) exhibit statistically significant agreement with the GT. In Figures~\ref{fig:SC25-ris-CR2240}, \& \ref{fig:SC25-ris-CR2266}, we present the CH estimations for the rising phase of Solar Cycle 25. During the rising phase, we see that coronal holes are getting more scarce, as more and more active regions emerge. We see that we can recover both very large and connected, and very small and isolated CHs. And also from the quantitative perspective, the results shown in Figures~\ref{fig:SC25-ris-CR2240} (CR2240), and ~\ref{fig:SC25-ris-CR2266} (CR2266) statistically agree well with the GT as calculated from Hotelling's $T^2$ test in Table~\ref{tab:sc25_Hotelling} (lines 3 and 4). In Figures~\ref{fig:SC25-max-CR2279} \& \ref{fig:SC25-max-CR2287}, we present the CH estimations for the maximum phase of Solar Cycle 25. As expected, CHs at maximum activity are more scarce and isolated. However, there are also more filaments which could mislead the model. We see that POP-CORN manages to identify the correct CH structures without mistaking filaments for CHs. And also, based on the statistical test presented in Table~\ref{tab:sc25_Hotelling}, we clearly see that the results shown in Figures~\ref{fig:SC25-max-CR2279} (CR2279), and ~\ref{fig:SC25-max-CR2287} (CR2287) statistically agree well with the GT as calculated from the Hotelling's $T^2$ test in Table~\ref{tab:sc25_Hotelling} (lines 5 and 6).

Overall, we observe that the CH estimations for Solar Cycle 25 from POP-CORN agree well with the GT, both qualitatively and quantitatively.

\begin{table}[ht]
\centering
\caption{Results of Hotelling's $T^{2}$ test comparing coronal hole (CH) identification from POP-CORN and ground truth (GT) for Solar Cycle 25 (Figure~\ref{fig:SC25-all-phases}).}
\label{tab:sc25_Hotelling}
\small
\setlength{\tabcolsep}{3.3pt}
\begin{tabular}{l l c c c c}
\hline
\textbf{Phase} & \textbf{CR} & $\boldsymbol{T^{2}}$ & \textbf{p} & \textbf{Decision} & \textbf{Result} \\
\hline
\multirow{2}{*}{Minimum} 
 & 2225 & 5.72 & 0.06 & Fail to reject $H_0$ & Stat. Significant \\
 & 2227 & 0.01 & 0.99 & Fail to reject $H_0$ & Stat. Significant \\
\hline
\multirow{2}{*}{Rising} 
 & 2240 & 0.27 & 0.87 & Fail to reject $H_0$ & Stat. Significant \\
 & 2266 & 0.69 & 0.72 & Fail to reject $H_0$ & Stat. Significant \\
\hline
\multirow{2}{*}{Maximum} 
 & 2279 & 3.82 & 0.16 & Fail to reject $H_0$ & Stat. Significant \\
 & 2287 & 0.18 & 0.92 & Fail to reject $H_0$ & Stat. Significant \\
\hline
\end{tabular}
\end{table}

\subsection{Solar Cycle 24}
\label{subsec:solar_cycle_24}

\begin{figure*}[ht]
\centering
\begin{subfigure}[b]{0.23\textwidth}
    \centering
    \includegraphics[width=\linewidth]{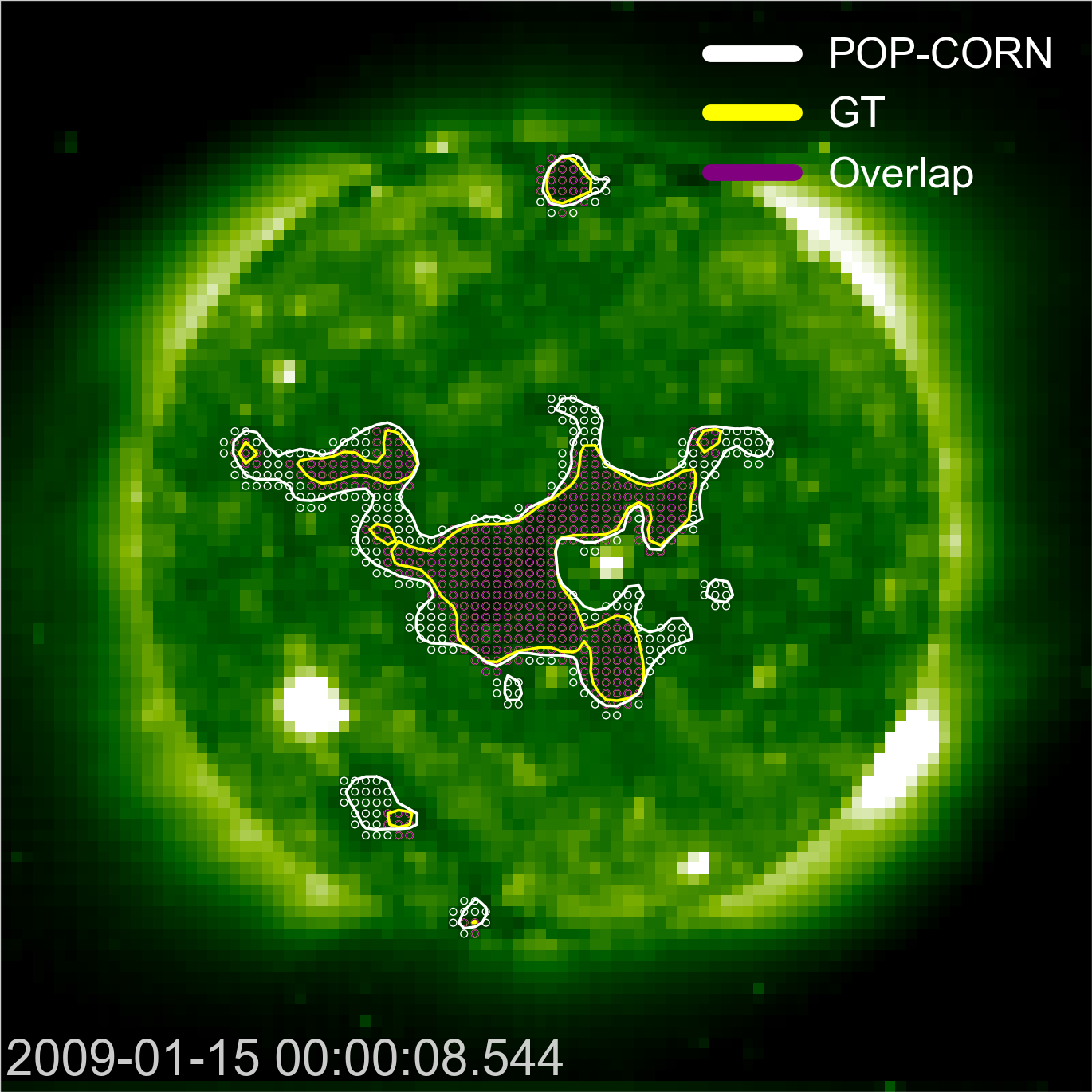}
    \caption{CR2079 (Minimum)}
    \label{fig:SC24-min-CR2079}
\end{subfigure}
\begin{subfigure}[b]{0.23\textwidth}
    \centering
    \includegraphics[width=\linewidth]{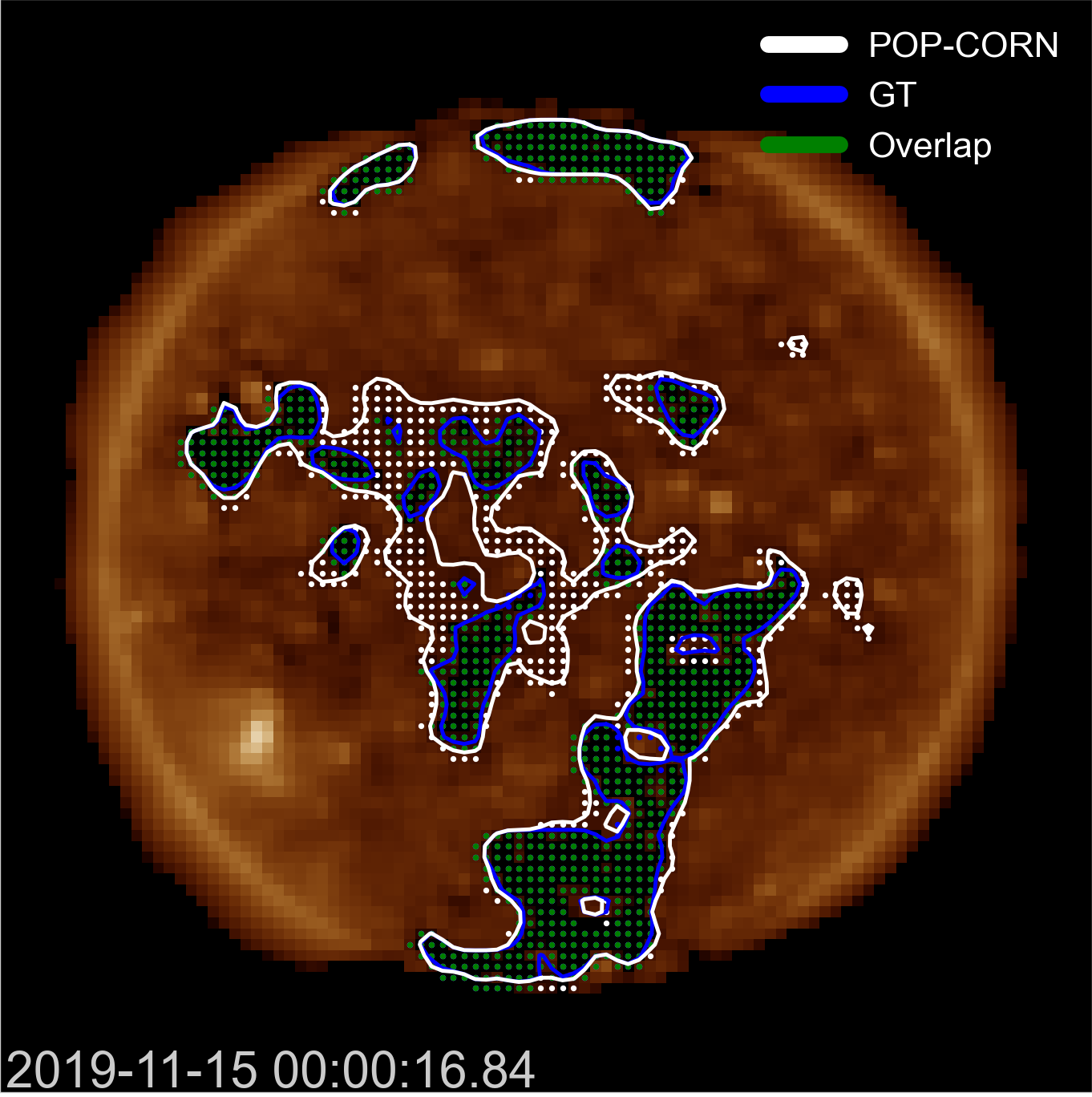}
    \caption{CR2224 (Minimum)}
    \label{fig:SC24-min-CR2224}
\end{subfigure}
\begin{subfigure}[b]{0.23\textwidth}
    \centering
    \includegraphics[width=\linewidth]{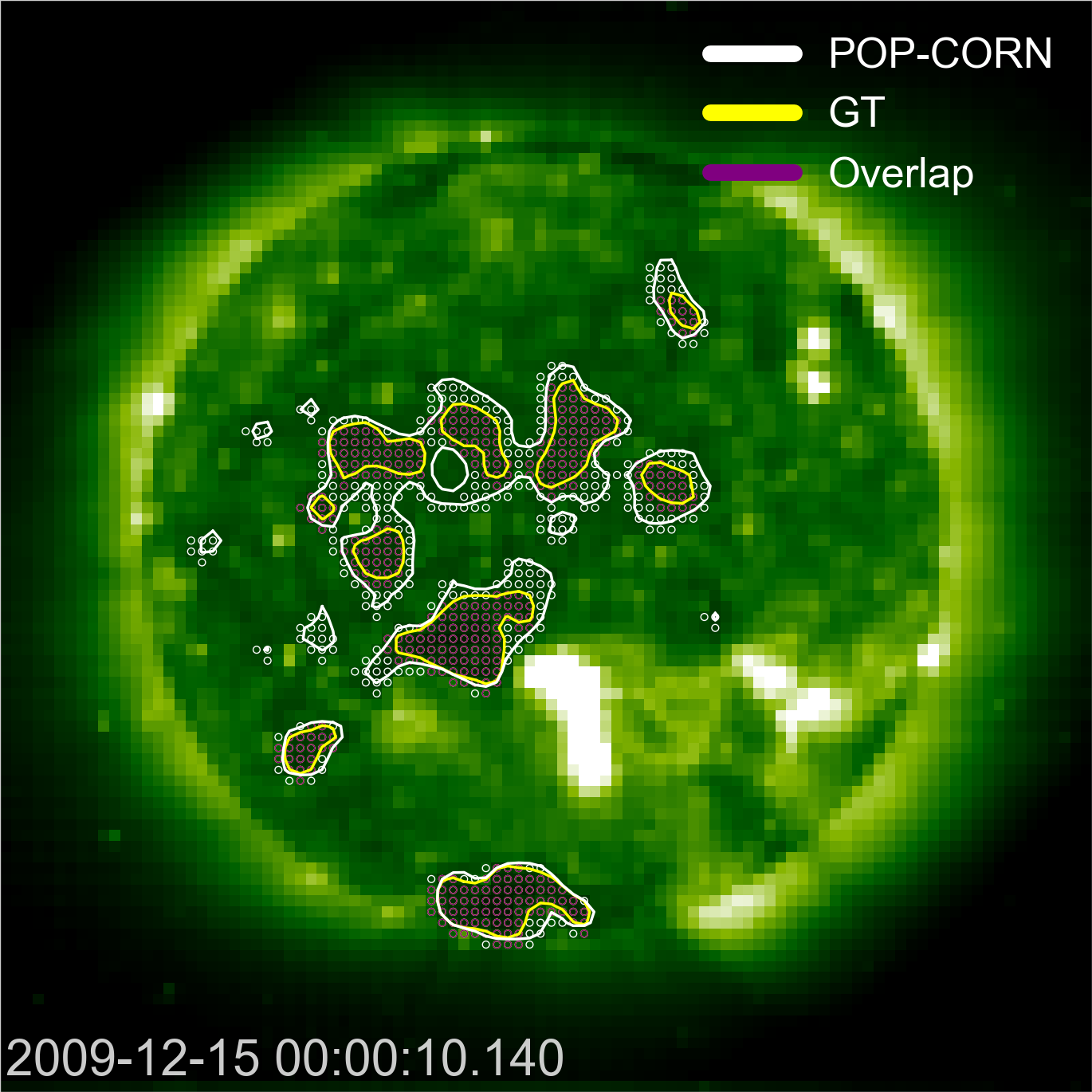}
    \caption{CR2091 (Rising)}
    \label{fig:SC24-ris-CR2091}
\end{subfigure}
\begin{subfigure}[b]{0.23\textwidth}
    \centering
    \includegraphics[width=\linewidth]{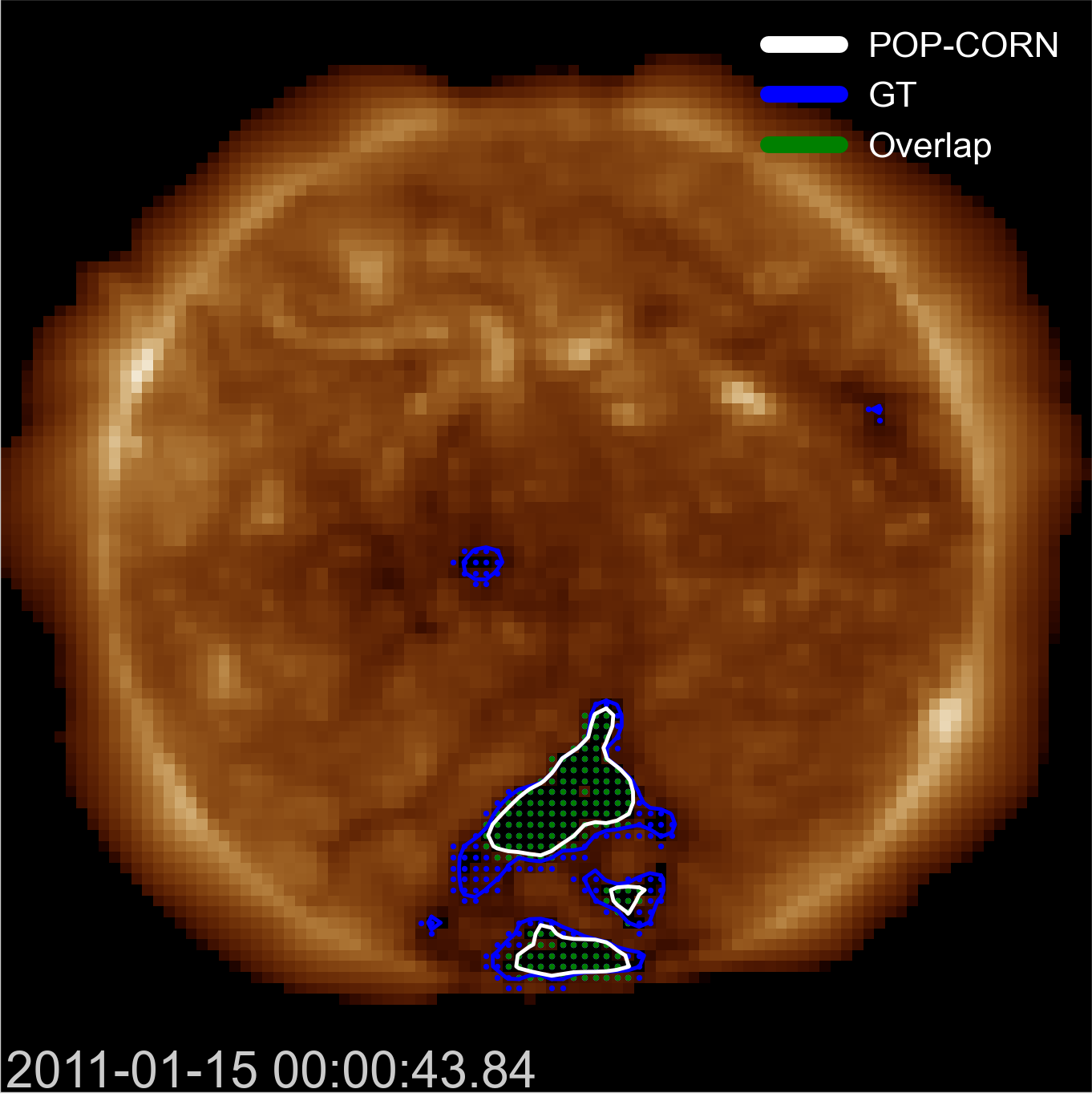}
    \caption{CR2105 (Rising)}
    \label{fig:SC24-ris-CR2105}
\end{subfigure}
\vspace{0.4em}
\begin{subfigure}[b]{0.23\textwidth}
    \centering
    \includegraphics[width=\linewidth]{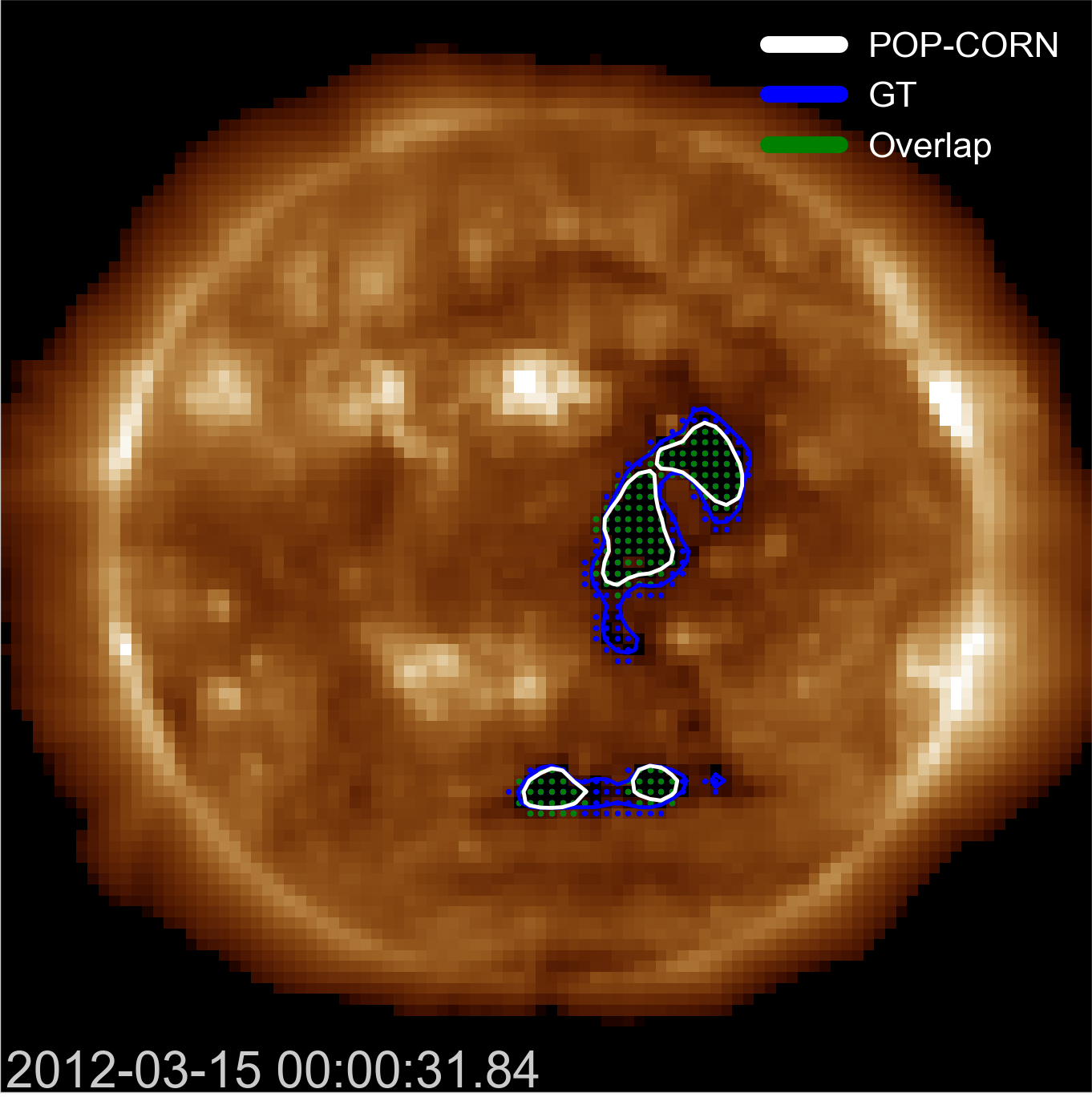}
    \caption{CR2121 (Maximum)}
    \label{fig:SC24-max-CR2121}
\end{subfigure}
\begin{subfigure}[b]{0.23\textwidth}
    \centering
    \includegraphics[width=\linewidth]{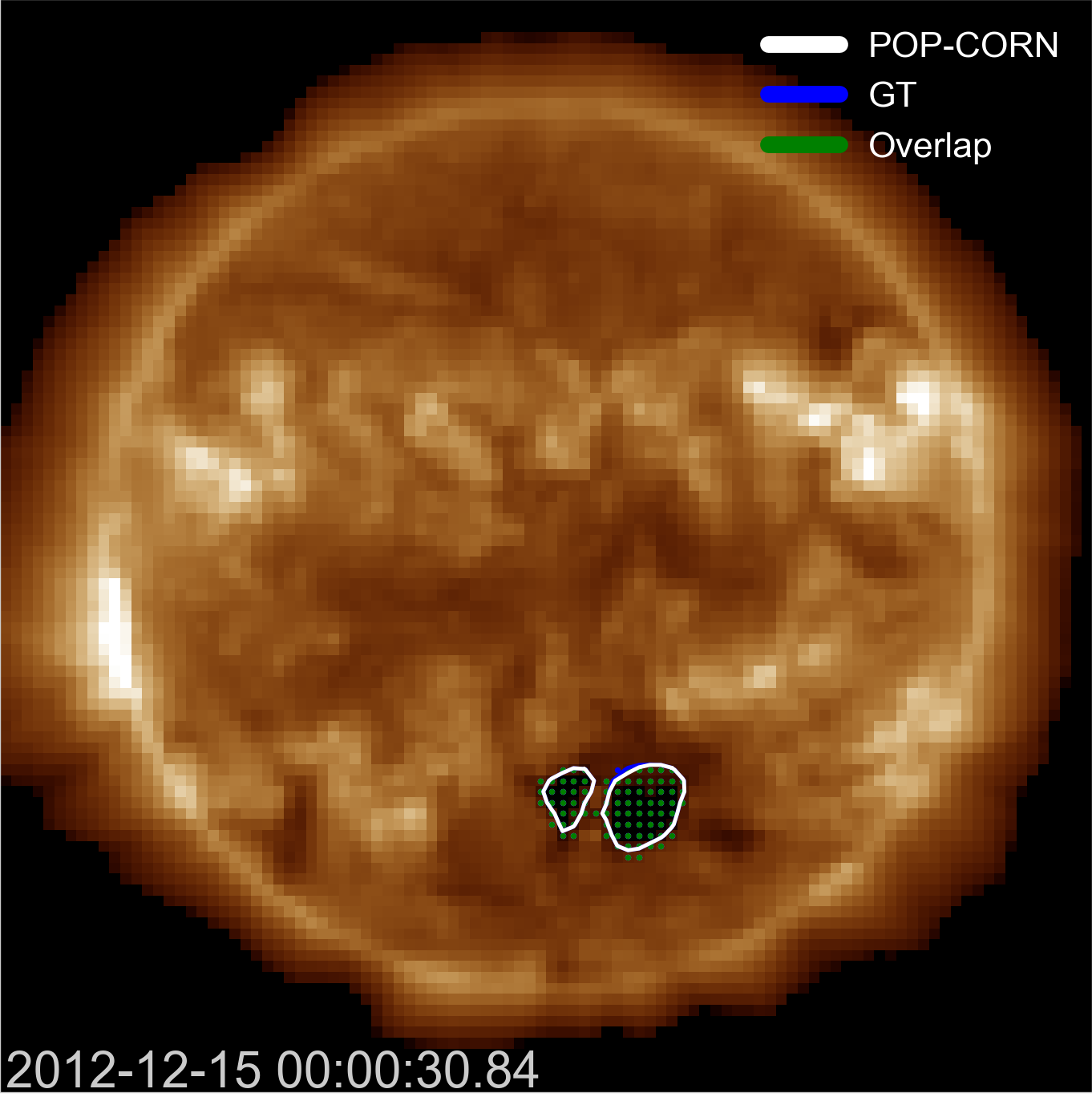}
    \caption{CR2131 (Maximum)}
    \label{fig:SC24-max-CR2131}
\end{subfigure}
\begin{subfigure}[b]{0.23\textwidth}
    \centering
    \includegraphics[width=\linewidth]{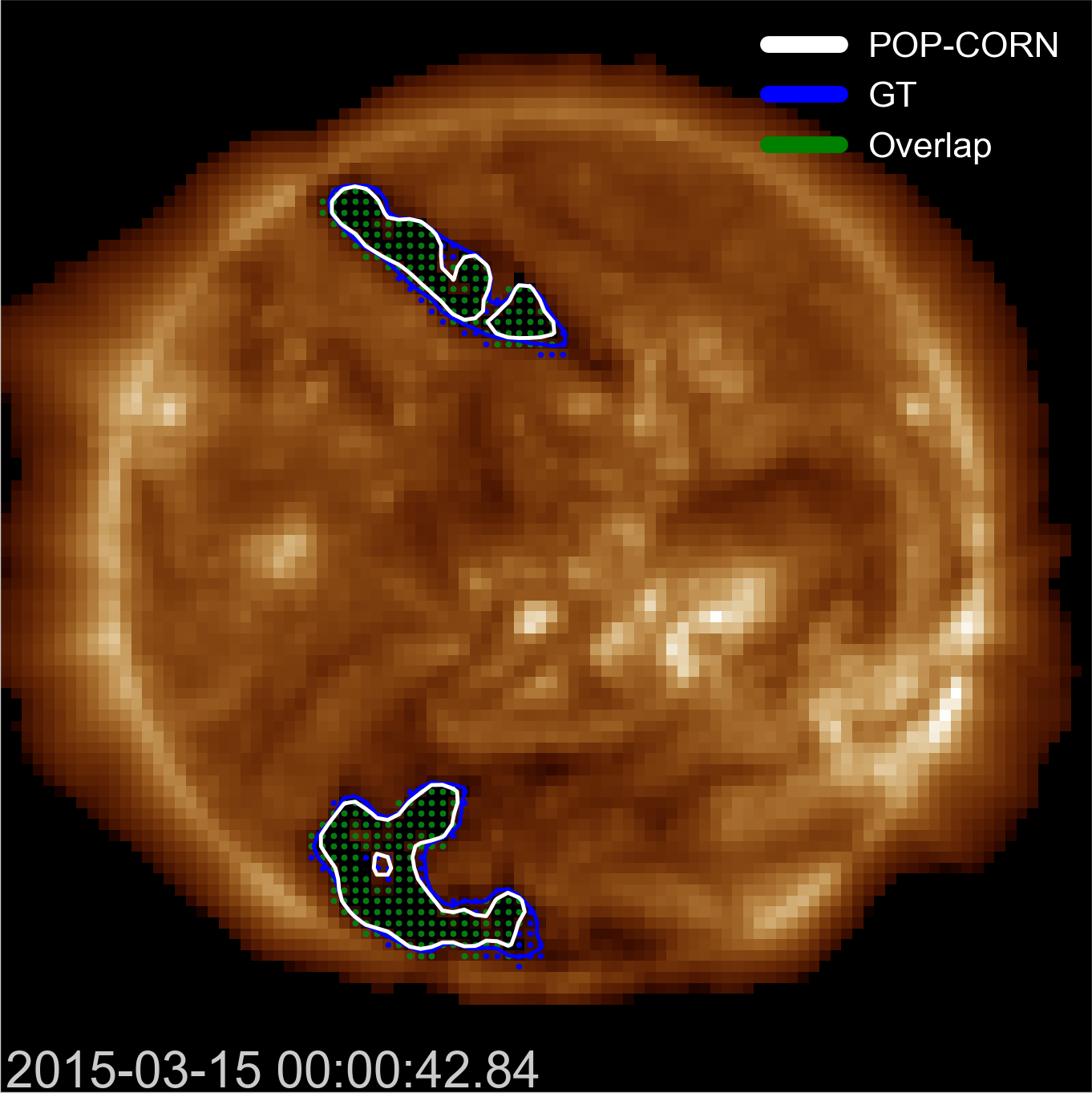}
    \caption{CR2161 (Declining)}
    \label{fig:SC24-dec-CR2161}
\end{subfigure}
\begin{subfigure}[b]{0.23\textwidth}
    \centering
    \includegraphics[width=\linewidth]{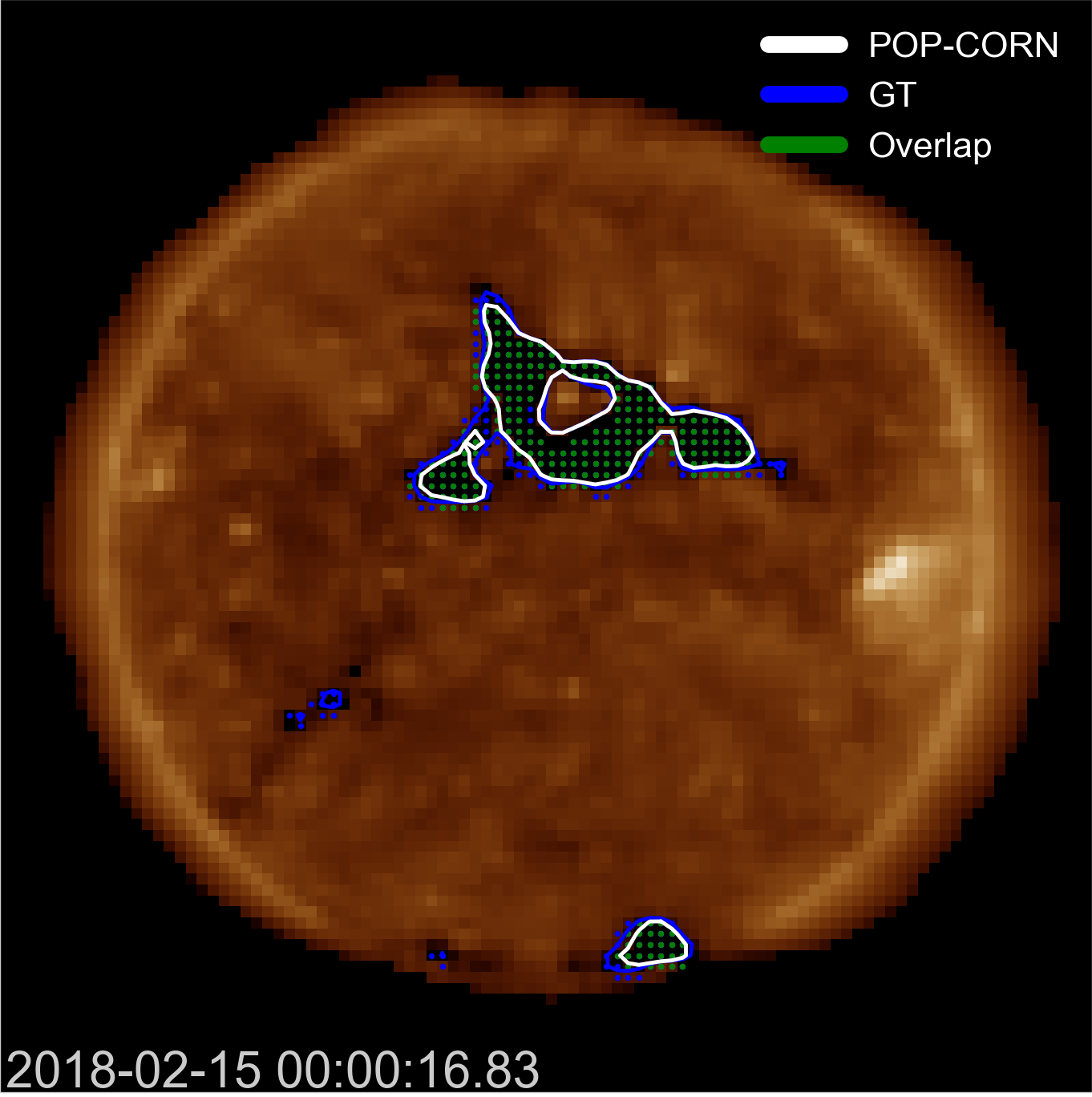}
    \caption{CR2200 (Declining)}
    \label{fig:SC24-dec-CR2200}
\end{subfigure}

\caption{Predicted coronal hole (CH) segmentation across different phases of Solar Cycle 24 (minimum, rising, maximum, and declining) using EUV images from SoHO/EIT at 195$\AA$ (\ref{fig:SC24-min-CR2079},\ref{fig:SC24-ris-CR2091}), and SDO/AIA at 193$\AA$ (\ref{fig:SC24-min-CR2224},\ref{fig:SC24-ris-CR2105},\ref{fig:SC24-max-CR2121},\ref{fig:SC24-max-CR2131},\ref{fig:SC24-dec-CR2161},\ref{fig:SC24-dec-CR2200}).}
\label{fig:SC24-all-phases}
\end{figure*}

The following images (Figure~\ref{fig:SC24-all-phases}) represent the CH predictions from POP-CORN for the solar minimum, rising, solar maximum, and the declining phases compared with the pixel contours from the GT. Here, we observe that the model can detect CH boundaries in each phase, both when the Sun is active and at minimum. From a qualitative perspective, we see that the model performs better across all phases of the solar cycle. 

Here, the Figure~\ref{fig:SC24-min-CR2079} (CR2079) illustrates the predictions from POP-CORN for the time \textbf{2009-01-15 00:00:08.544}, and here we apply the SDO-SoHO threshold conversion. We see, qualitatively and from the statistical test (Table~\ref{tab:sc24_Hotelling}), that the estimated CH contours in Figures~\ref{fig:SC24-min-CR2079} (CR2079) \& ~\ref{fig:SC24-min-CR2224} (CR2224) agree well with the GT. Here, for the solar minimum phase, as discussed for Solar Cycle 25 (Subsection~\ref{subsec:solar_cycle_25}), we also observe low contrast in the pictures due to the absence of active regions. Here, we observe good agreement between the GT and POP-CORN contours. In Figures~\ref{fig:SC24-ris-CR2091} \& \ref{fig:SC24-ris-CR2105}, we present the CH estimations for the rising phase of Solar Cycle 24. During the rising phase of Solar Cycle 24, the coronal hole at lower latitudes and the equator begin to disappear, and more and more active regions emerge. In this case, too, we see that we can recover both very large and connected CHs, as well as very small and isolated CHs. The statistical test (Table~\ref{tab:sc24_Hotelling}) shows that the predictions for Figures~\ref{fig:SC24-ris-CR2091} (CR2091) and \ref{fig:SC24-ris-CR2105} (CR2105) agree well with the GT. Here again, for Figure~\ref{fig:SC24-ris-CR2091} (CR2091), we use the SDO-SoHO threshold conversion. In Figures~\ref{fig:SC24-max-CR2121}, \& \ref{fig:SC24-max-CR2131}, we present the CH estimations for the maximum phase of Solar Cycle 24. As expected, for the solar maximum case in Solar Cycle 25 (Subsection \ref{subsec:solar_cycle_25}), we observe that POP-CORN identifies the correct CH structures without mistaking filaments for CHs. The statistical test (Table~\ref{tab:sc24_Hotelling}) shows that the predictions for Figures~\ref{fig:SC24-max-CR2121} (CR2121) and \ref{fig:SC24-max-CR2131} (CR2131) agree well with the GT. The Figures~\ref{fig:SC24-dec-CR2161}, \& \ref{fig:SC24-dec-CR2200}, illustrates the CH contour comparison for the declining phase of Solar Cycle 24. Here, we see that the POP-CORN estimates the CHs well, in agreement with the GT. At the beginning of the declining phase, we observe filaments and active regions appearing at the mid and lower latitudes, and still, POP-CORN captures the CHs well both qualitatively (Figure~\ref{fig:SC24-dec-CR2161} CR2161) and statistically (Table~\ref{tab:sc24_Hotelling} - line 7). At the end of the declining phase we observe less number of active regions. Here, dark regions appear at the mid latitude, and the POP-CORN is able to recover only the CH from other dark regions well quantitatively (Figure~\ref{fig:SC24-dec-CR2200} CR2200) and statistically (Table~\ref{tab:sc24_Hotelling} - line 8).     

\begin{table}[ht]
\centering
\caption{Results of Hotelling's $T^{2}$ test comparing coronal hole (CH) identification from POP-CORN and ground truth (GT) for Solar Cycle 24 (Figure~\ref{fig:SC24-all-phases}).}
\label{tab:sc24_Hotelling}
\small
\setlength{\tabcolsep}{3.5pt}
\begin{tabular}{l l c c c c}
\hline
\textbf{Phase} & \textbf{CR} & $\boldsymbol{T^{2}}$ & \textbf{p} & \textbf{Decision} & \textbf{Result} \\
\hline
\multirow{2}{*}{Minimum} 
 & 2079 & 2.06 & 0.36 & Fail to reject $H_0$ & Stat. Significant \\
 & 2224 & 3.09 & 0.21 & Fail to reject $H_0$ & Stat. Significant \\
\hline
\multirow{2}{*}{Rising} 
 & 2091 & 5.54 & 0.06 & Fail to reject $H_0$ & Stat. Significant \\
 & 2105 & 4.22 & 0.12 & Fail to reject $H_0$ & Stat. Significant \\
\hline
\multirow{2}{*}{Maximum} 
 & 2121 & 2.49 & 0.29 & Fail to reject $H_0$ & Stat. Significant \\
 & 2131 & 0 & 1.00 & Fail to reject $H_0$ & Stat. Significant \\
\hline
\multirow{2}{*}{Declining} 
 & 2161 & 0.89 & 0.64 & Fail to reject $H_0$ & Stat. Significant \\
 & 2200 & 5.50 & 0.07 & Fail to reject $H_0$ & Stat. Significant \\
\hline
\end{tabular}
\end{table}

\subsection{Solar Cycle 23}
\label{subsec:solar_cycle_23}

\begin{figure}[htpb!]
\centering
\begin{subfigure}[htpb]{0.45\linewidth}
    \centering
    \includegraphics[width=\linewidth]{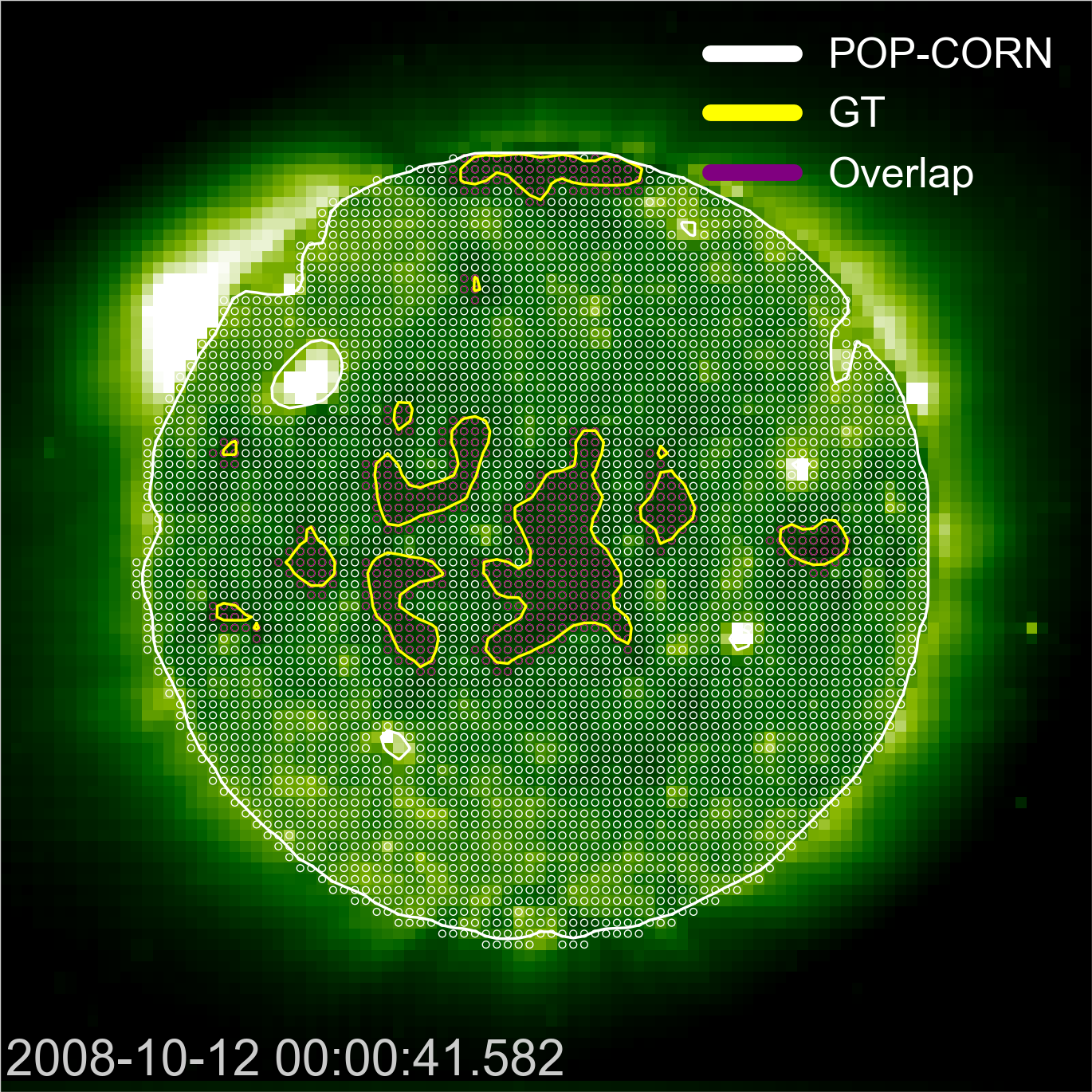}
    \caption{CR2075 (MIN)}
\end{subfigure}
\begin{subfigure}[htpb]{0.45\linewidth}
    \centering
    \includegraphics[width=\linewidth]{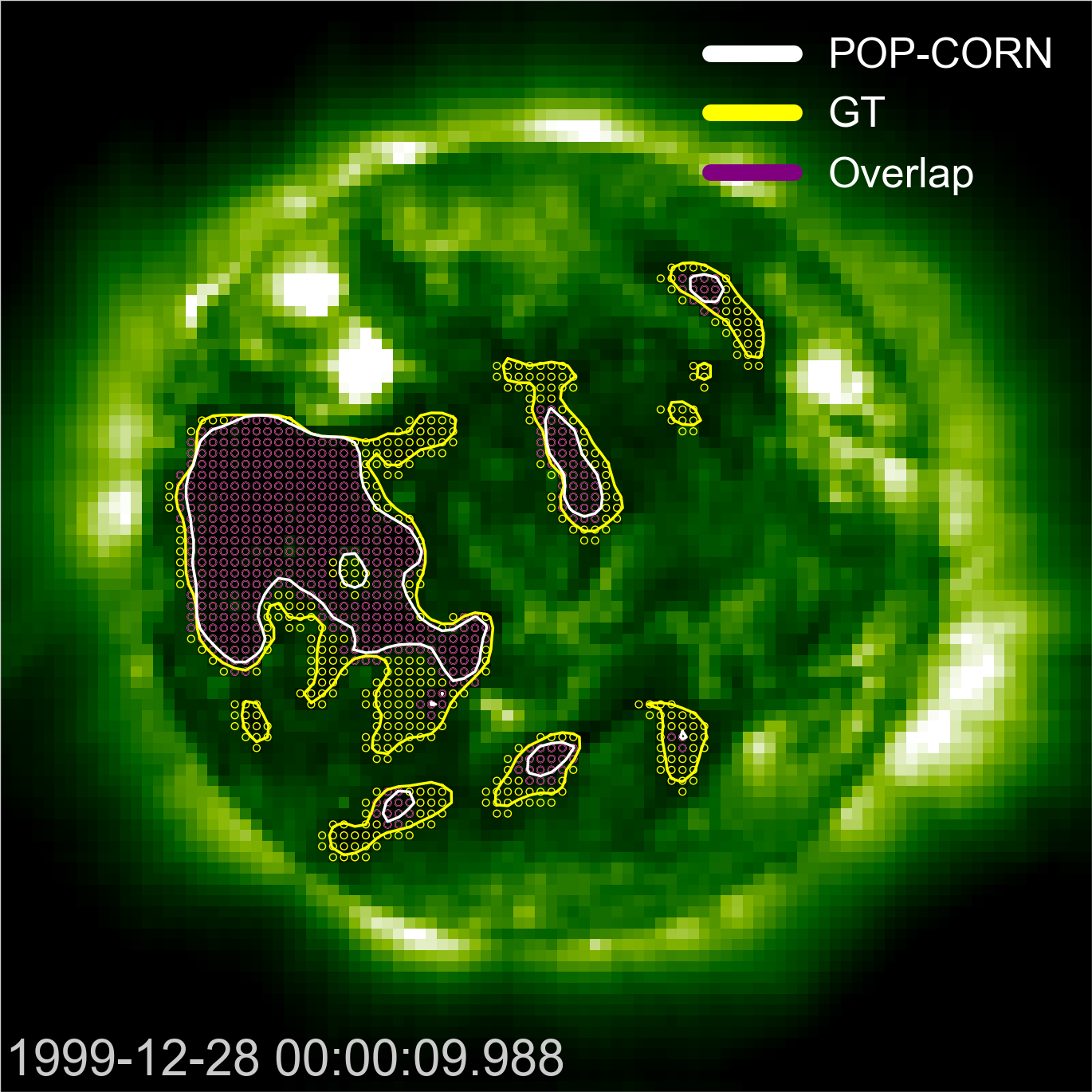}
    \caption{CR1957 (RISING)}
\end{subfigure}
\vspace{0.15cm}
\begin{subfigure}[htpb]{0.45\linewidth}
    \centering
    \includegraphics[width=\linewidth]{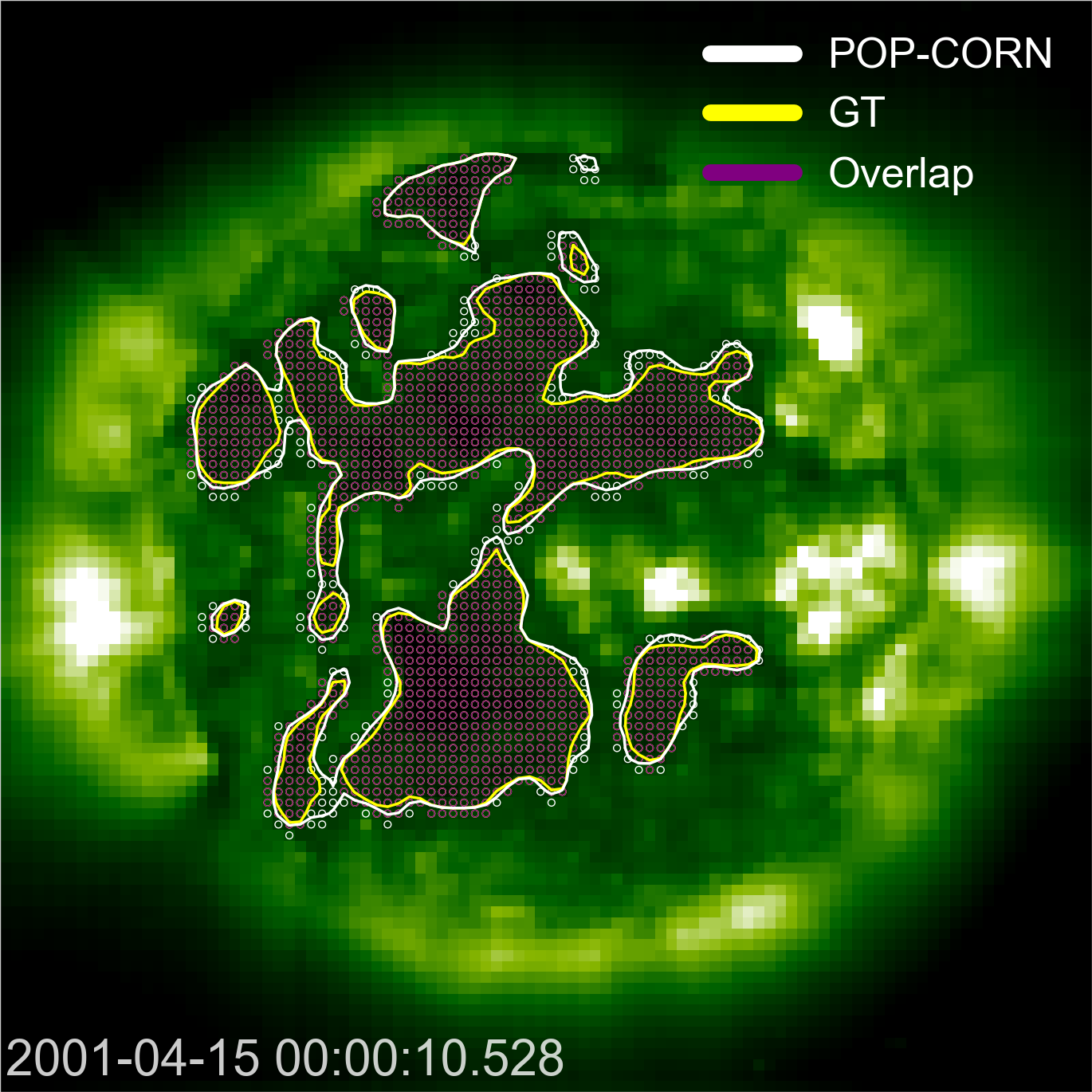}
    \caption{CR1975 (MAX)}
\end{subfigure}
\begin{subfigure}[htpb]{0.45\linewidth}
    \centering
    \includegraphics[width=\linewidth]{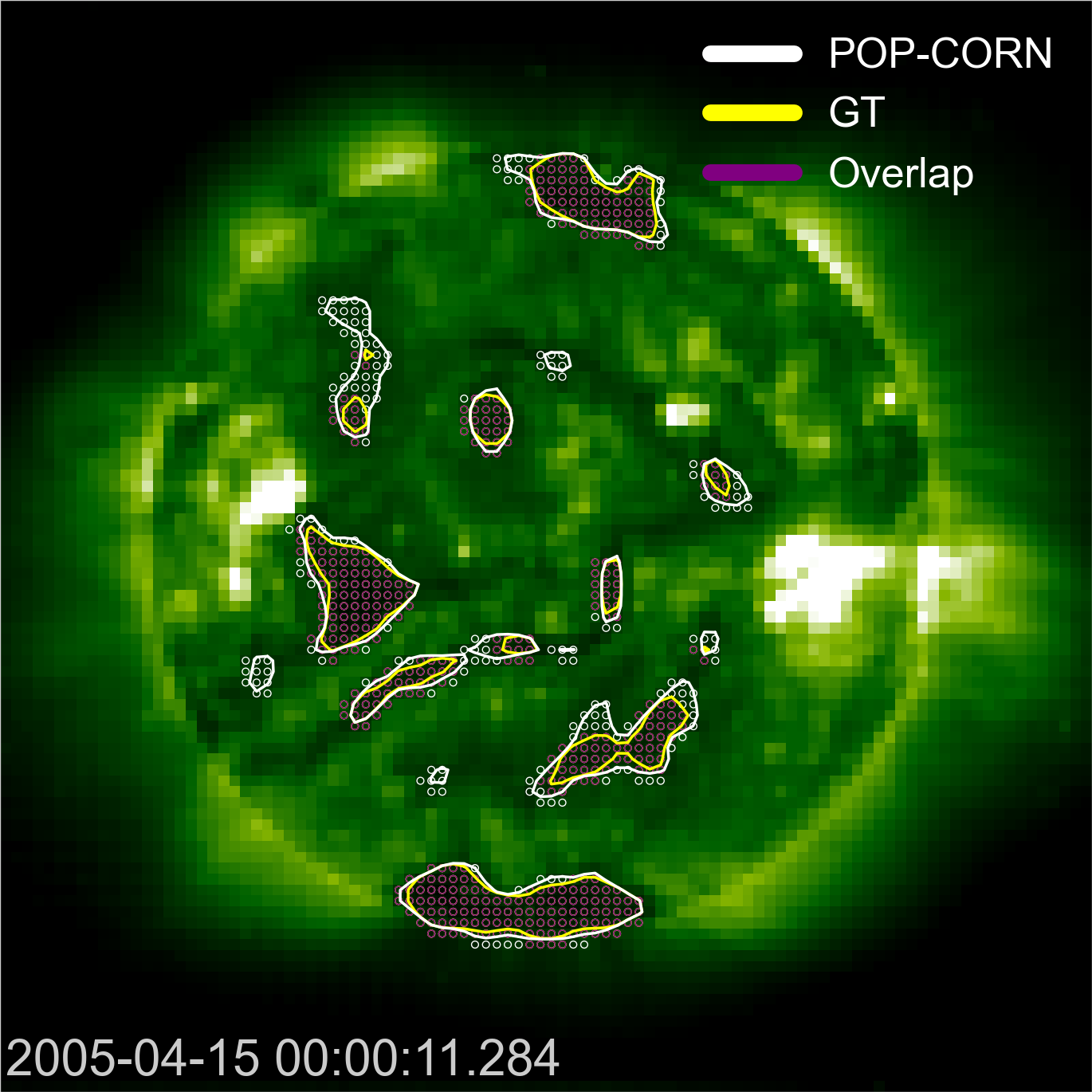}
    \caption{CR2028 (DECLINING)}
\end{subfigure}
\caption{Predicted CH segmentation for Solar Cycle 23 during the solar minimum, rising, maximum, and declining period for EUV images for SoHO/EIT 195$\AA$.}
\label{fig:SC23}
\end{figure}

The results below, shown in Figure~\ref{fig:SC23}, are for Solar Cycle 23, and we show our CH estimations for solar minimum, rising, maximum, and declining phases of this cycle. The comparison here is between the predictions from POP-CORN and the GT. The CH contours are filled with pixels in the SOHO/EIT images: the predicted CHs are shown in white, the GTs in yellow, and the overlapping regions in purple. Since the images used here are from SoHO/EIT, the threshold values obtained from the model are not 100\% accurate. We plan to improve predictions for Solar Cycle 23 in the future; therefore, we have included only one result for each phase in this cycle.

From the results, we see that the model estimates CHs well across the maximum and declining phases, but it tends to overestimate during the rising and solar minimum. This is why the picture at solar minimum appears white everywhere: in this specific case, the entire solar disk is selected as a CH contour, which is, of course, incorrect. This overestimation should be taken into account when improving the NN model for Solar Cycle 23 in the future. The main reason for the overestimation is the model's greater sensitivity to small threshold variations, as discussed in Section~\ref{subsec:Relation_between_the_threshold_values}, which leads to the inclusion of all darker regions as CH boundaries. Table~\ref{tab:sc23_hotelling's} presents the statistical hypothesis comparison, explaining how CH contour matching is performed; the hypotheses are described in detail in Appendix~\ref{sec:HT}. We see from Table~\ref{tab:sc23_hotelling's} that for the maximum (\textit{CR1975}) and declining (\textit{CR2028}) phases, the predictions agree well with the GT statistically. It is, however, less critical to have good results with SoHO/EIT than with SDO/AIA, as new observatories will have properties closer to SDO/AIA.

\begin{table}[ht]
\centering
\caption{Hotelling's $T^{2}$ test results comparing CH identification from POP-CORN and GT for Solar Cycle 23 (Figure~\ref{fig:SC23}).}
\label{tab:sc23_hotelling's}
\footnotesize
\setlength{\tabcolsep}{3.3pt}
\begin{tabular}{l l c c c c}
\hline
\textbf{CR} & \textbf{Phase} & $\boldsymbol{T^{2}}$ & \textbf{p} & \textbf{Decision} & \textbf{Result} \\
\hline
2075 & Minimum   & 14.05 & 0    & Reject $H_0$        & Not Stat. Significant \\
1957 & Rising    & 19.16 & 0    & Reject $H_0$        & Not Stat. Significant \\
1975 & Maximum   & 0.29  & 0.87 & Fail to reject $H_0$  & Stat. Significant \\
2028 & Declining & 2.34  & 0.31 & Fail to reject $H_0$  & Stat. Significant \\
\hline
\end{tabular}
\end{table}

\section{Discussion} 
\label{sec:discussion}

Unlike many other models described in \citet{Reiss_2024}, POP-CORN detects CHs using a threshold-based detection technique in Python image processing. The model is trained on data from solar cycles 23, and 24, and the user can obtain predictions for optimal threshold values to detect the CHs for SoHO/EIT in the wavelength of 195$\AA$, SDO/AIA in the wavelength of 193$\AA$, and for SDO/AIA composite images in the wavelengths of 171$\AA$ + 193$\AA$ + 211$\AA$, at any time during cycles 23-25. At present, the model performs better on cycles 24-25, and we are improving it to achieve more accurate predictions for cycle 23. A key difference from other NN models is that we do not train EUV images to capture the physics and properties of the solar cycle and predict the CH regions. 

\begin{figure*}[ht]
\centering
\begin{subfigure}[b]{0.30\textwidth}
    \centering
    \includegraphics[width=\linewidth]{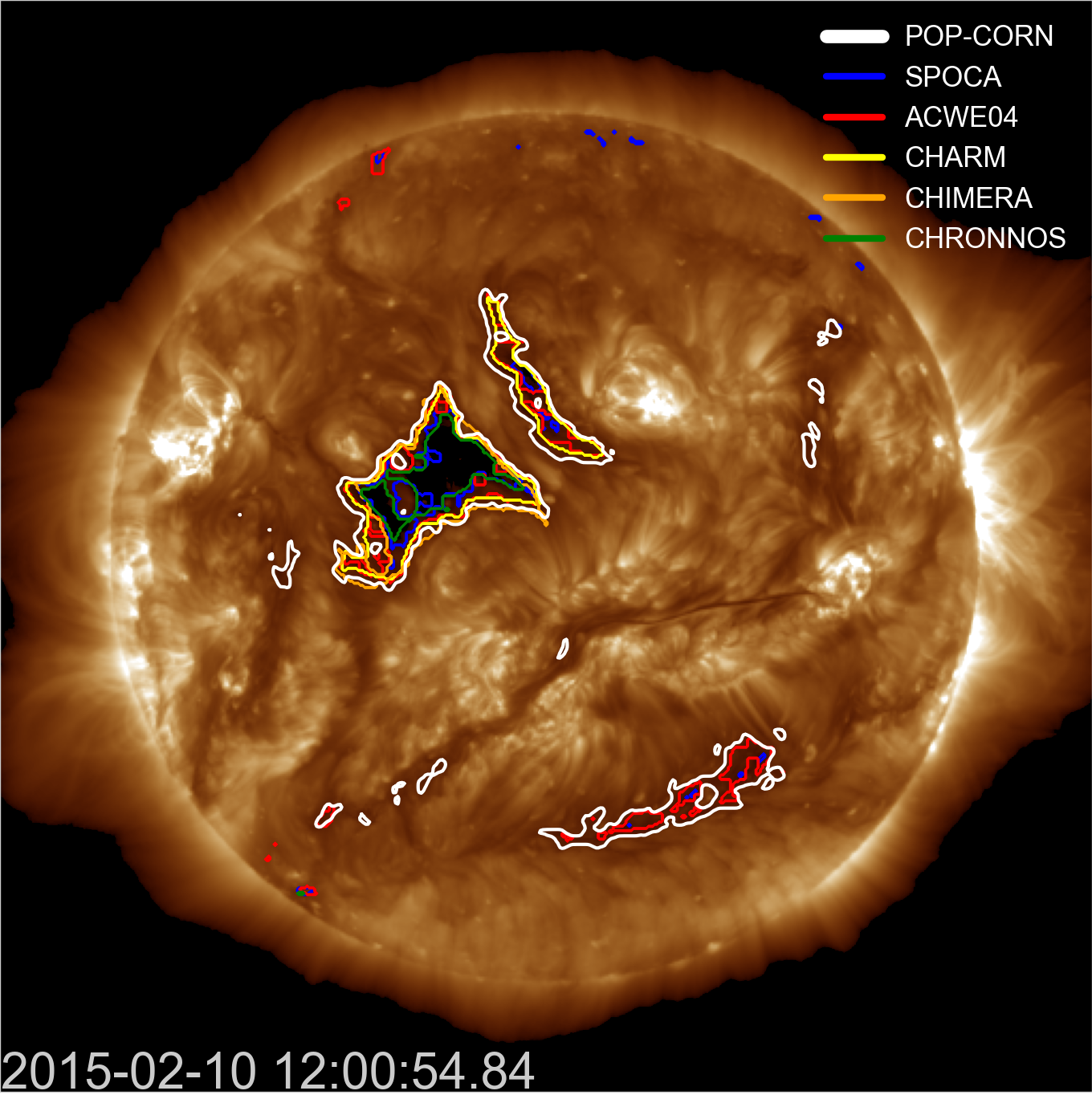}
    \caption{CR2160}
    \label{fig:CH-comp-CR2160}
\end{subfigure}
\begin{subfigure}[b]{0.30\textwidth}
    \centering
    \includegraphics[width=\linewidth]{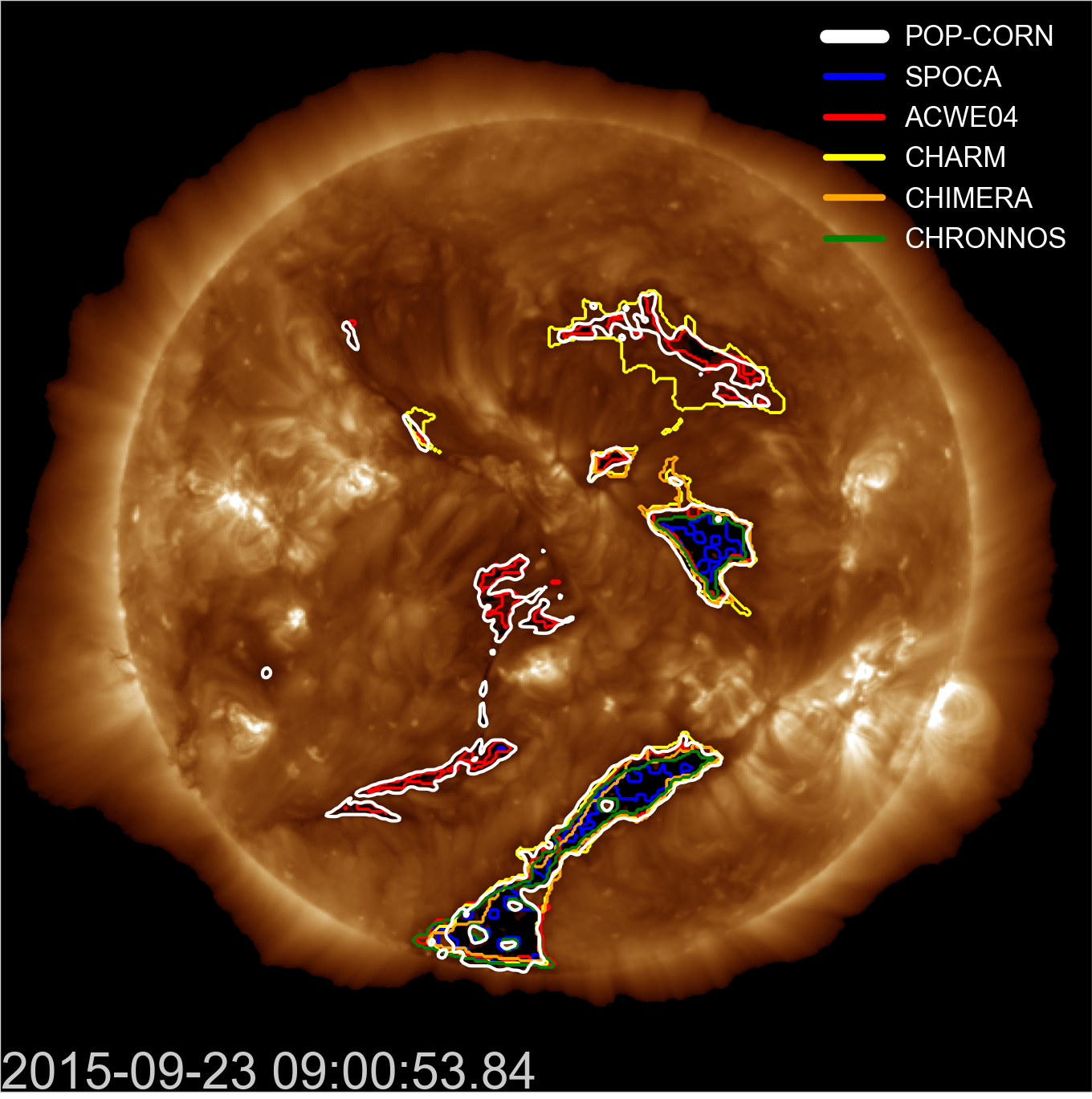}
    \caption{CR2168}
    \label{fig:CH-comp-CR2168}
\end{subfigure}
\begin{subfigure}[b]{0.30\textwidth}
    \centering
    \includegraphics[width=\linewidth]{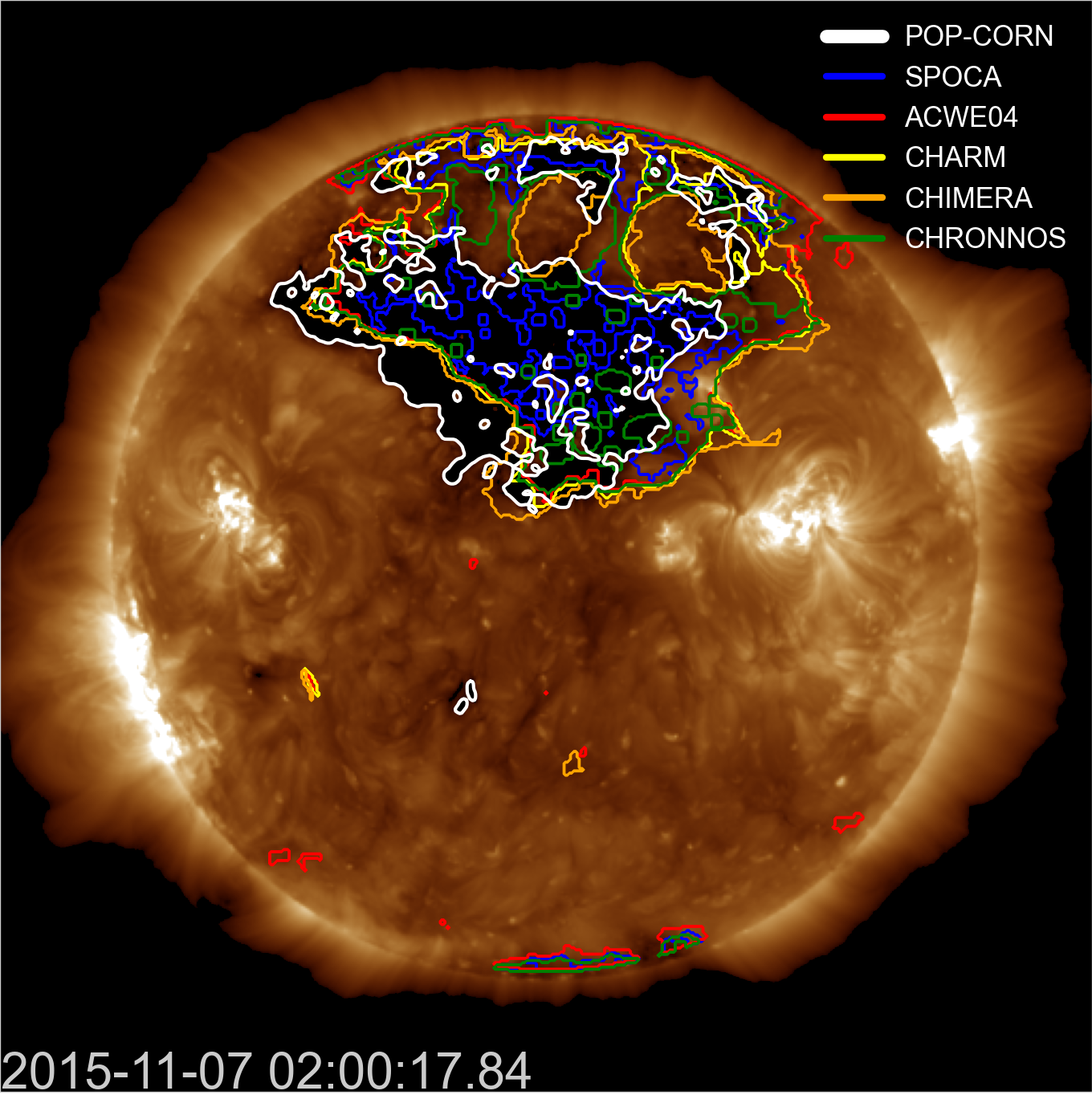}
    \caption{CR2170}
    \label{fig:CH-comp-CR2170}
\end{subfigure}
\vspace{0.5em}
\begin{subfigure}[b]{0.30\textwidth}
    \centering
    \includegraphics[width=\linewidth]{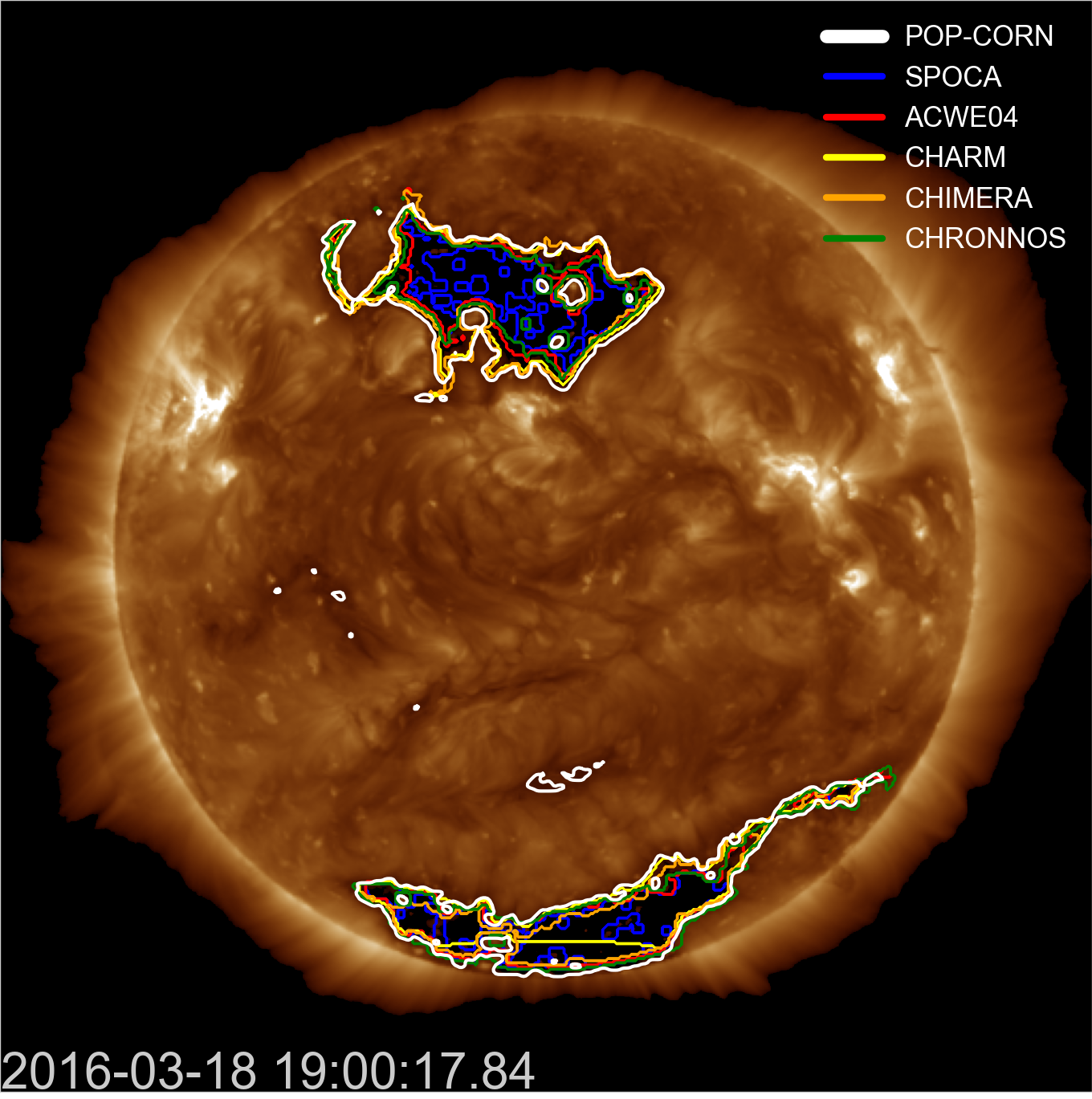}
    \caption{CR2175}
    \label{fig:CH-comp-CR2175}
\end{subfigure}
\begin{subfigure}[b]{0.30\textwidth}
    \centering
    \includegraphics[width=\linewidth]{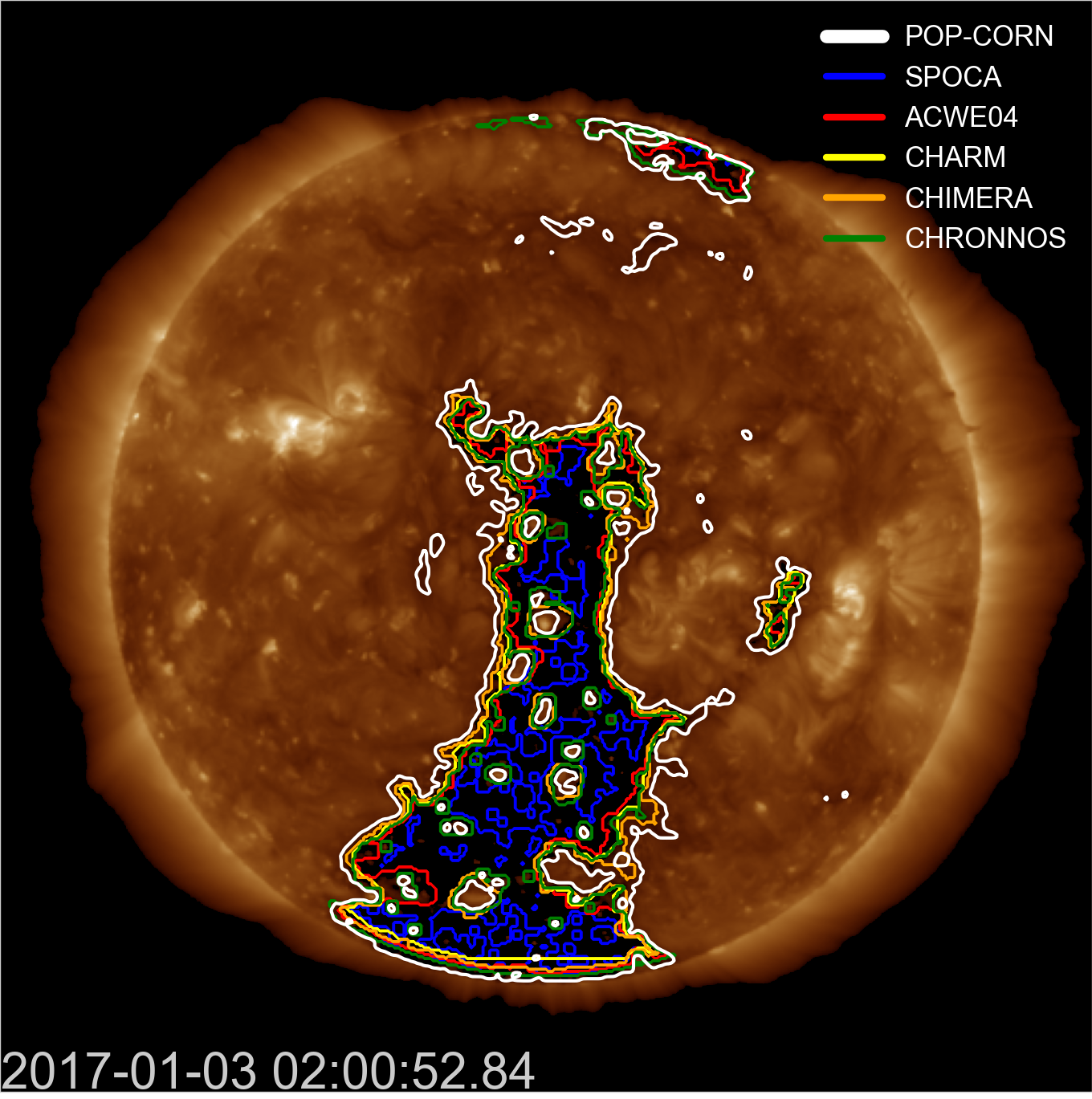}
    \caption{CR2185}
    \label{fig:CH-comp-CR2185}
\end{subfigure}
\begin{subfigure}[b]{0.30\textwidth}
    \centering
    \includegraphics[width=\linewidth]{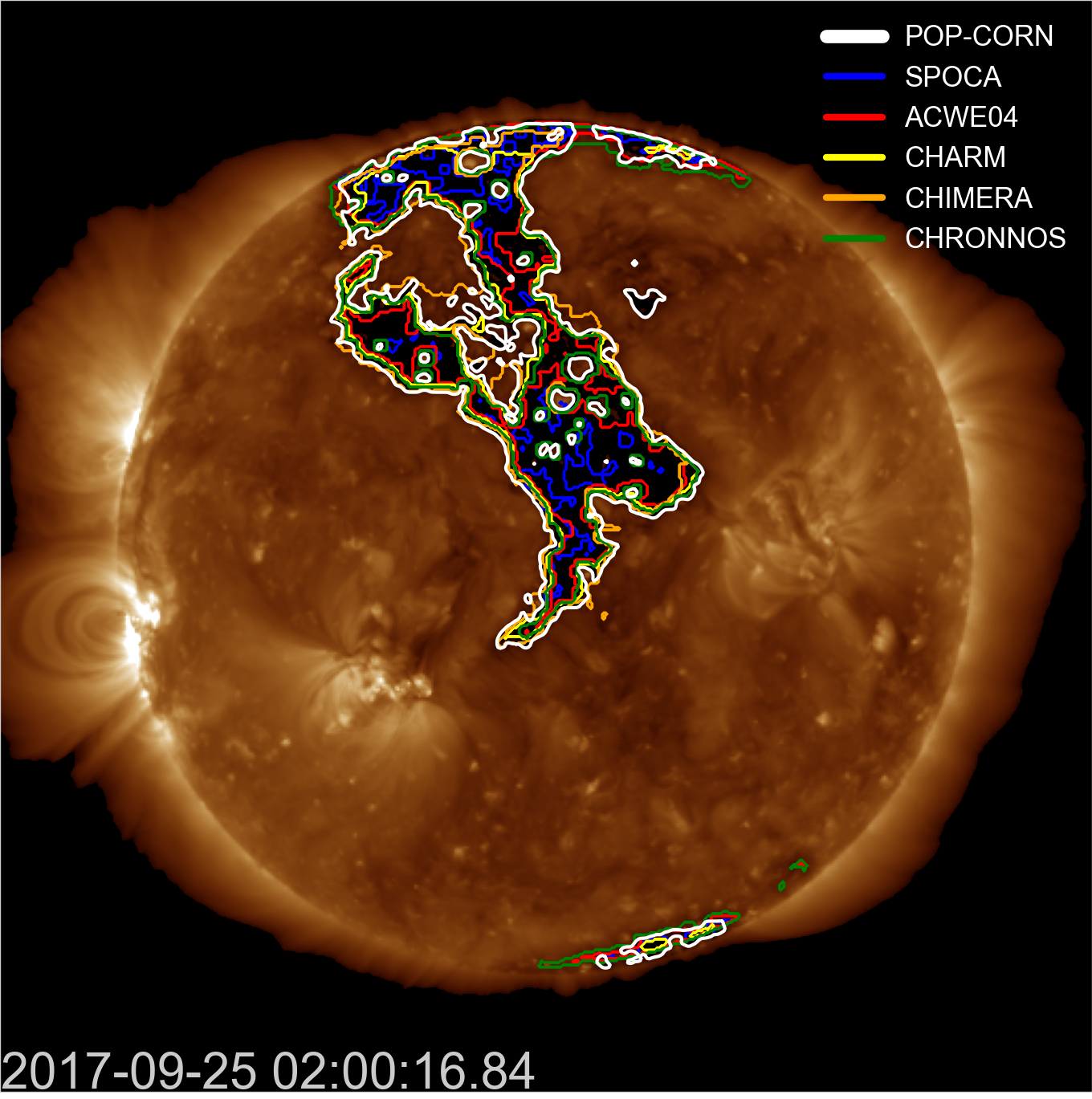}
    \caption{CR2195}
    \label{fig:CH-comp-CR2195}
\end{subfigure}
\caption{Comparison of the POP-CORN estimations with existing coronal hole detection tools: SPOCA, ACWE04, CHARM, CHIMERA, and CHRONNOS for EUV images from SDO/AIA 193$\AA$ \citet{Reiss_2024}.}
\label{fig:CH_comparison}
\end{figure*}

Here, we also present a qualitative comparison of the CH contours reproduced by POP-CORN which is illustrated in Figure~\ref{fig:CH_comparison} with those from some of the models discuss in \citet{Reiss_2024}: the selected models are \textit{SPOCA} (\citet{ACWE}), \textit{ACWE} (\citet{ACWE}), \textit{CHARM} (\citet{CHARM}), \textit{CHIMERA} (\citet{CHIMERA}), and \textit{CHRONNOS} (\citet{CHRONNOS}). For more details on the models above, see the references. We display the CH contours for POP-CORN in cyan, \textit{SPOCA} in blue, \textit{ACWE04} in red, \textit{CHARM} in yellow, \textit{CHIMERA} in orange, and \textit{CHRONNOS} in green. Here, we present CH contour comparisons for POP-CORN for the dates \textit{2015-02-10 12:00:54.84, 2015-09-23 09:00:53.84, 2015-11-07 02:00:17.84, 2016-03-18 19:00:17.84, 2017-01-03 02:00:52.84,} and \textit{2017-09-25 02:00:16.84} to compare with the aforementioned CH detection tools (refer to Figure~\ref{fig:CH_comparison}). Here, we see that from the comparison, the POP-CORN estimations agree well qualitatively with the CH detection tools mentioned above.

From the qualitative overview shown in the results Section~\ref{sec:results}, POP-CORN estimates CHs, as the other models do (refer to Figure~\ref{fig:CH_comparison}), and in some cases agrees better than they do. Specifically, for the case shown in Figure~\ref{fig:CH-comp-CR2170} (CR2170), we clearly see that the POP-CORN model estimates CH contours well, capturing CHs by location, and it also captures small CH contours at mid latitudes, without overestimating filaments and other dark regions. The improved performance of our model is due to training on only the most important features of the large-scale structures of the solar corona; this approach avoids additional artifacts in image intensities and other errors, such as corrupted images. The corrupted images are mainly due to image noise arising from solar energetic particles, scattered light, and missing blocks caused by unstable EIT telemetry in the SoHO/EIT images \citep{Hamada_2019}. To compare the pixel contour coordinates, it's important to select an appropriate statistical test to validate the results. Therefore, we use Hotelling's $T^{2}$, statistical test (\citep{Aparisi1996}, \citep{ChenHsieh2007}) to compare the CH pixel coordinates within the contours from the GT and the POP-CORN. Here, the GT defines the CHs detected from a threshold value observed in the eye when viewing the EUV image. The statistical comparisons are made for the CH predictions illustrated in Section~\ref{sec:model} for solar cycles 23, 24, and 25. Please refer to Appendix~\ref{sec:HT} for equations and the hypothesis definitions for Hotelling's $T^2$ statistical test. 

\begin{figure}[htpb]
\centering
\begin{subfigure}{0.48\linewidth}
    \centering
    \includegraphics[width=\linewidth]{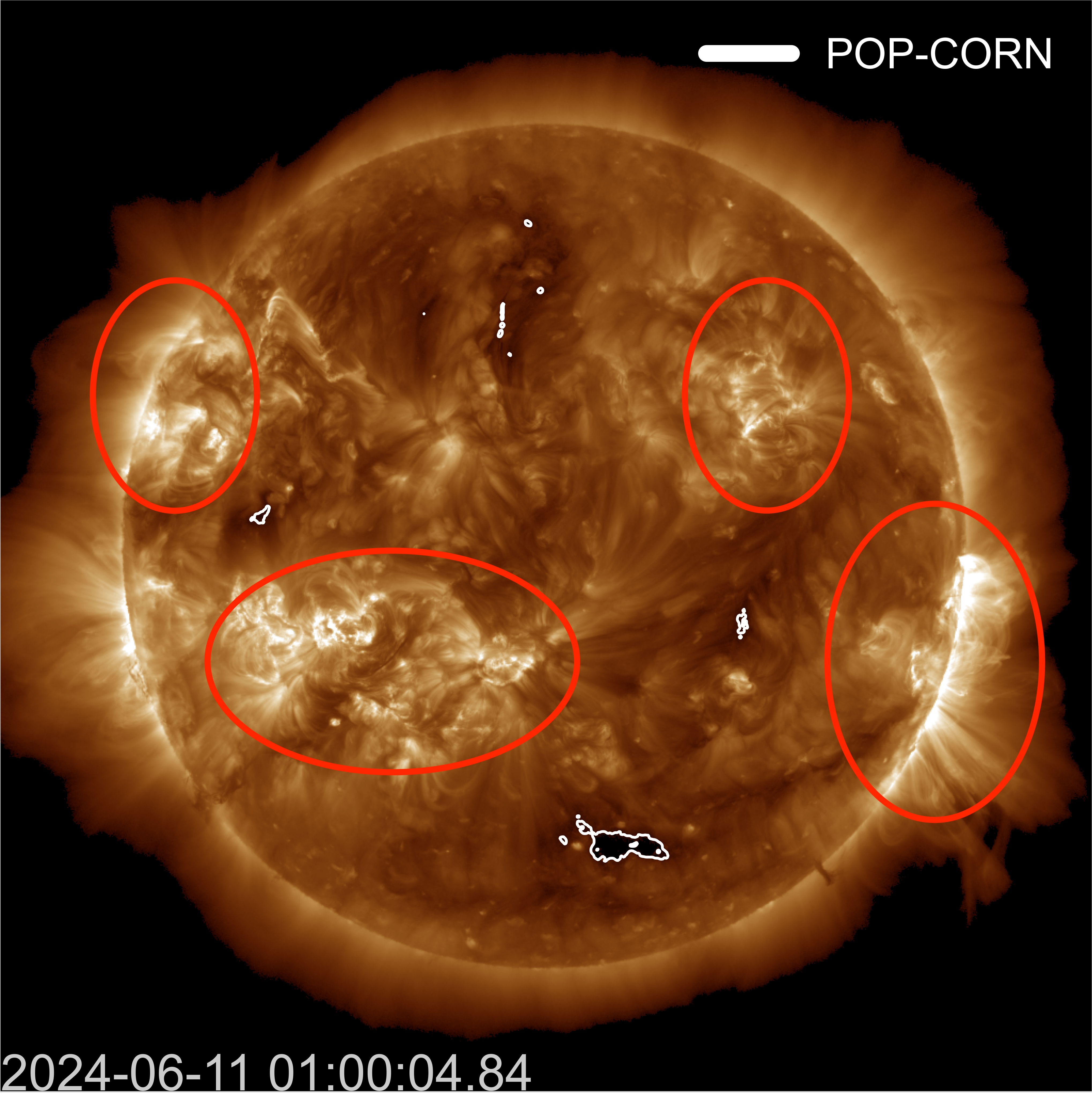}
\end{subfigure}
\begin{subfigure}{0.48\linewidth}
    \centering
    \includegraphics[width=\linewidth]{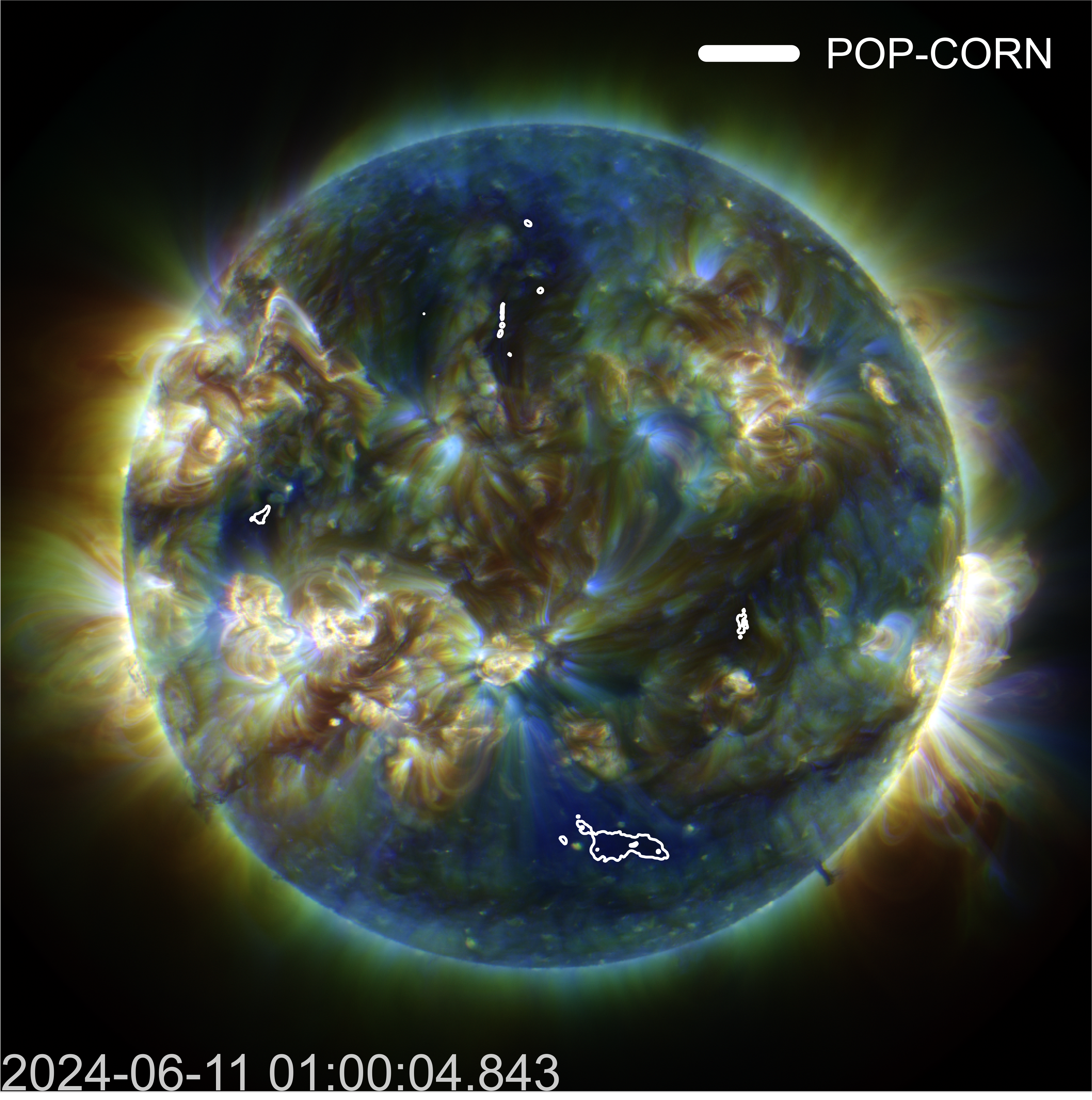}
\end{subfigure}
\caption{Coronal hole predictions obtained from POP-CORN on \textbf{\textit{2024-06-11 01:00:04.843}}. The locations of solar flares are indicated inside red circles. On the left is the SDO-AIA \textit{193{$\AA$}}, and on the right is the composite images (\textit{171{$\AA$}} + \textit{193{$\AA$}} + \textit{211{$\AA$}})}
\label{fig:X-Class_flare}
\end{figure}

Another key factor is how feature engineering and the inputs of the NN model influence model training. POP-CORN takes categorical (binary) inputs that describe the properties and spatial distribution of large-scale structures in the SC, as well as the phase of the solar cycle. This approach is unique and provides an efficient way to encode the physics of the main structures of the SC into the NN model using categorical representations. From the predictions shown in Figure~\ref{fig:X-Class_flare}, we observe that when the Sun is active or when we observe a large number of solar flares occurring within the LOS, it is challenging to identify the dark regions or the CHs in our case, because of the brightness of the image due to the higher intensity. The flare regions are circled in red in Figure~\ref{fig:X-Class_flare}, and our model can capture the CH regions quite well, as illustrated on the \textit{2024-06-11 01:00:04.843}. The predicted threshold is comparably low here, and since the NN model incorporates flare class and location during training, we obtain a reliable threshold for detecting CHs and identify CH regions with good qualitative agreement. In the same image (Figure~\ref{fig:X-Class_flare}), the right panel illustrates the composite image for the same LOS with the wavelength of (\textit{171{$\AA$}} + \textit{193{$\AA$}} + \textit{211{$\AA$}}). The composite image provides a clearer view of the CHs than SDO/AIA \textit{193{$\AA$}}.  

The evolution of the CHs of the solar cycle is another key effect that we consider during the image feature processing, where we categorized the CH morphology change into the four phases of solar cycles, 23-25. When the Sun is at minimum, we observe many dark regions and may sometimes find it challenging to see the CHs, but when the Sun is active, it's quite easy to distinguish them. However, the case is different when many flares and CMEs are present, as discussed in the previous paragraph. Therefore, we included the phase change into the model training (solar minimum, rising, maximum, and declining phases). From the results presented in Section~\ref{sec:results}, we find that these phases capture the full range of solar activity and that using them as input variables yields strong predictive performance for CHs across solar cycles 23, 24, and 25. 

Another effect that we consider is the magnetic classification of the sunspot groups. Different sunspot groups exhibit varying active-region strengths, which affect the brightness or intensity of the LOS-EUV image and are important factors in identifying the threshold value for CH segmentations. The importance of manually incorporating these properties into the model is that it can better capture small-scale CHs than models such as Convolutional Neural Networks (CNNs). We can also identify significant regions, such as coronal holes, which is a main advantage of POP-CORN over other models. We can also capture short-term events, such as changes in EUV image intensity after a CME or during coronal dimming. Therefore, these properties are important for training the model, since CH detection relies on a threshold-based approach. 

The results presented in this work have strong implications for how CH detection affects the properties of large-scale structures in the solar corona. The brightness of EUV images varies with temperature, plasma density, and magnetic field strength in this region. The strong magnetic field implies greater heating and leads to bright EUV emission, which directly affects the threshold value used in the Python image processing discussed in Section~\ref{subsec:Identifying_the_CH_boundaries}. Here, the NN model (POP-CORN) provides optimal threshold values based on the locations and intensities of large-scale structures, which directly affect the brightness of the solar corona. We see that the class of solar flares and the different magnetic activity of active regions greatly affect the intensities of the EUV images, thereby influencing the threshold value used to predict the CH contours. POP-CORN is a data-driven numerical model that incorporates the physics of the solar corona as data. We implemented the features from the large-scale structures of the solar corona (refer to Section~\ref{subsec:feature_eng}) that we believe are going to be more important in detecting the CH regions, and from the results (refer to Section~\ref{sec:results}), we compared them with the GT and other CH detection methods. 


\section{Conclusions}
\label{sec:Conclusions}

In this study, we developed a CH detection tool (POP-CORN) based on large-scale structures of the solar corona by categorizing those features into phase, class, and spatial distribution. As discussed in Section~\ref{subsec:feature_eng}, the input of the model is binary (0/1) based on all the categorical features, and the output is the threshold value, which needs to identify the CH contours either from SoHO/EIT 195$\AA$, or from SDO/AIA 193$\AA$. When we train the model, we evaluate accuracy on both the training and validation data. 25$\%$ of the data is used to validate the model, and we consider it accurate based on the accuracy score and the prevention of overfitting and underfitting (refer to Section~\ref{subsec:Model_Validation}).   
As a summary, we see that POP-CORN performs well, capturing CHs during solar cycles 24 and 25. We are still working towards improving the SDO-SOHO conversion to accurately predict CHs for solar cycle 23. From the results, we observe that when the Sun is active, and at a minimum, the model performs well compared to the GT. We conclude that the main reason for this is that the most important features for determining the CH regions are the properties and locations of the large-scale structures of the SC, including active regions, solar flares, coronal mass ejections (CMEs), and filaments. Therefore, we identify these structures as vital for identifying CH regions in threshold-based detection schemes. We confirm that manually incorporating these features, rather than training on EUV images across solar cycles, is more effective in detecting CHs, both qualitatively and quantitatively. We observe that, in the results, the POP-CORN estimates CH regions far better when the LOS of the Sun passes through many active or flare regions, where it is difficult to distinguish CH regions from other dark regions, such as filament structures.

For the CH predictions in Section~\ref{sec:results} for different solar cycles 25, 24, and 23, we make the comparison quantitative by using the GT threshold to compare with the POP-CORN predictions, and then perform statistical comparisons to validate the results. From the statistical test, we confirmed that in most cases, the CH predictions are statistically agreeing with the GT, except for a few cases as observed in Table~\ref{tab:sc25_Hotelling}, Table~\ref{tab:sc24_Hotelling}, and Table~\ref{tab:sc23_hotelling's} for CR2075, and CR1957, we observe that the CH contours from the NN model is not agreeing with the GT statistically. But in the remaining cases, Hotelling's $T^{2}$ is low (<~10), and the p-value is > 0.05. (fail to reject the null-hypotheses ($H_o$), or the POP-CORN results are statistically significant). The main reason for the disagreement is that the SDO-SoHO conversion is underperforming and needs improvement. As discussed in the model section (refer to Section~\ref{subsec:Relation_between_the_threshold_values}), the reason is that the different pixel scales in the SDO/AIA (0.6 arcsec/pixel) and in SOHO/EIT (2.6 or 5.26 arcsec/pixel) lead to greater sensitivity in SOHO/EIT to small threshold variations.

In the future, we plan to improve the SDO-SOHO conversion, with a strong model to couple with the main CH detection pipeline, where it can overcome the greater sensitivity to small threshold variations. The NN model in POP-CORN is trained to obtain threshold values for SDO/AIA 193$\AA$ and SoHO/EIT 195$\AA$, with an image resolution of 1024. Finally, we plan to integrate the POP-CORN CH detection tool into a pipeline for validating solar wind models. This work is part of the WindTRUST project, and to achieve the project's goal, we plan to develop a validation tool for the WindPredict polytropic (WP-$\gamma$) model \citet{Reville(2017)}. This validation tool is based on coronal hole observations (from POP-CORN), but additional diagnostics will be added for complementary assessment, including the current sheet, magnetic connectivity, and UV and white-light emissions. Ultimately, we plan to provide a quantitative score based on the WP-$\gamma$ predictions.

\begin{acknowledgements}
  This work received funding from the French National Research Agency under reference ANR-23-ASZC-0004 (ANR Astrid WindTRUST).
  This work used data provided by the MEDOC data and operations centre (CNES / CNRS / Univ. Paris-Saclay), \href{http://medoc.ias.u-psud.fr/}{http://medoc.ias.u-psud.fr/}, and FIDO data \citep{sunpy}.
  Courtesy of NASA/SDO and the AIA, EVE, and HMI science teams.
  SOHO is a project of international cooperation between ESA and NASA.
  The NN model is trained on data obtained from the HEK database for the large-scale structures of the SC. \href{https://www.lmsal.com/hek/api.html}{https://www.lmsal.com/hek/api.html}.
\end{acknowledgements}


\appendix
\label{sec:appendix}

\section{Additional Figures}
\label{sec:Additional_figures}

\begin{figure}[htbp]
    \centering
    \includegraphics[width=0.35\textwidth]{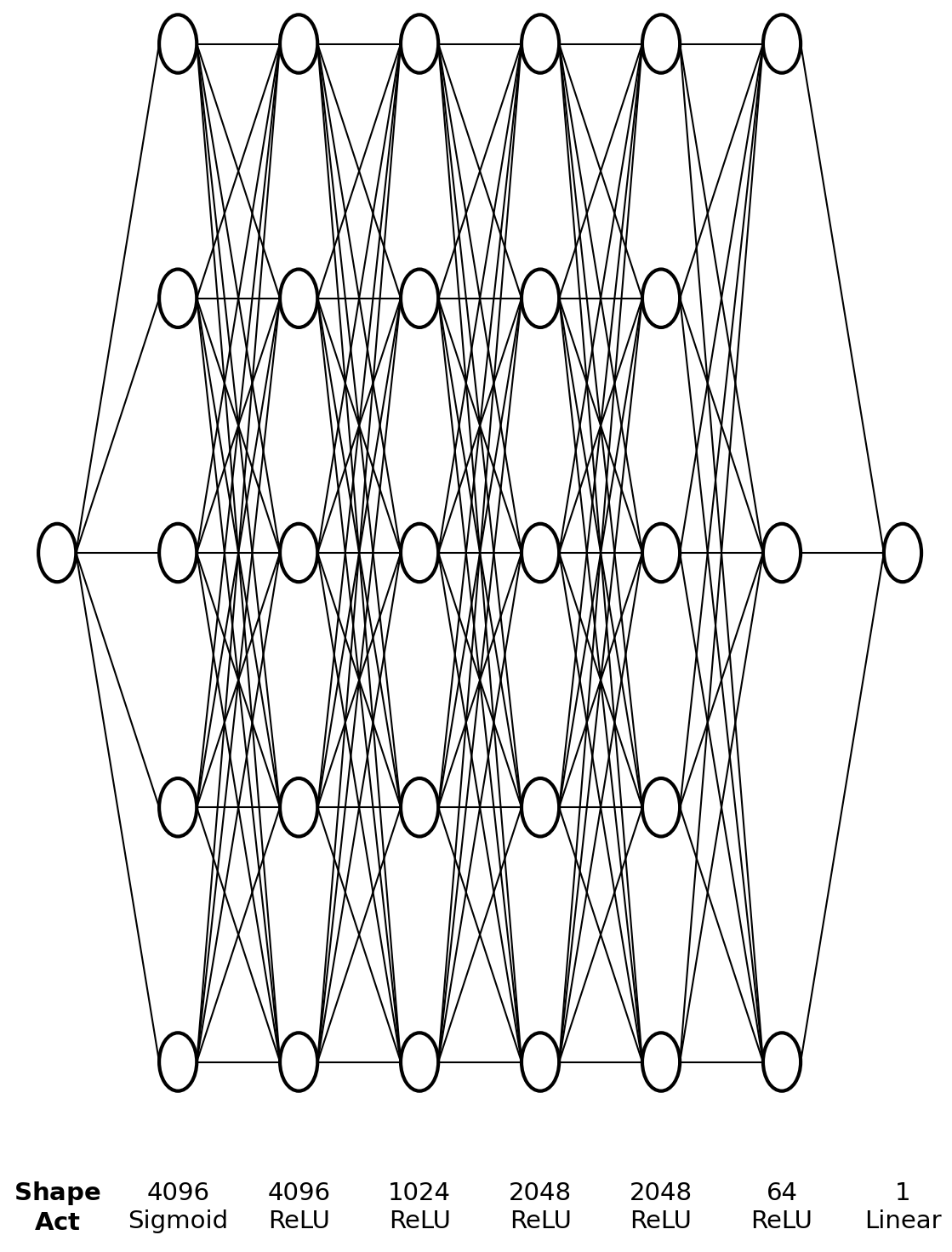}
    \caption{Neural network architecture illustrating layer dimensions and activation functions.}
    \label{fig:NN_Config}
\end{figure}
\section{Hotelling’s $T^{2}$ Test}
\label{sec:HT}

Here, Hotelling's $T^{2}$ statistic measures the multivariate distance between the means, and the p-value defines whether the difference is statistically significant. Here, we set $\alpha$ = 0.05 and consider the results statistically significant if $p > 0.05$.

\begin{align}
T^2 =
\frac{n_1 n_2}{n_1 + n_2}
(\bar{\mathbf{X}} - \bar{\mathbf{Y}})^\top
\mathbf{S}_p^{-1}
(\bar{\mathbf{X}} - \bar{\mathbf{Y}})
\label{eq:hotelling_t2}
\end{align}

The sample mean vectors are defined as
\begin{align}
\bar{\mathbf{X}}
&= \frac{1}{n_1}\sum_{i=1}^{n_1}\mathbf{X}_i,
\label{eq:mean_X} \\[4pt]
\bar{\mathbf{Y}}
&= \frac{1}{n_2}\sum_{i=1}^{n_2}\mathbf{Y}_i,
\label{eq:mean_Y}
\end{align}

and the pooled covariance matrix is given by
\begin{align}
\mathbf{S}_p
&= \frac{(n_1 - 1)\mathbf{S}_1 + (n_2 - 1)\mathbf{S}_2}{n_1 + n_2 - 2},
\label{eq:pooled_cov}
\end{align}
where the individual sample covariance matrices are
\begin{align}
\mathbf{S}_1
&= \frac{1}{n_1 - 1}\sum_{i=1}^{n_1}
(\mathbf{X}_i - \bar{\mathbf{X}})
(\mathbf{X}_i - \bar{\mathbf{X}})^\top,
\label{eq:sample_cov_X} \\[6pt]
\mathbf{S}_2
&= \frac{1}{n_2 - 1}\sum_{i=1}^{n_2}
(\mathbf{Y}_i - \bar{\mathbf{Y}})
(\mathbf{Y}_i - \bar{\mathbf{Y}})^\top.
\label{eq:sample_cov_Y}
\end{align}

\noindent\textbf{where:}
\begin{align*}
\mathbf{S}_p & : \text{pooled covariance matrix}, \\
\mathbf{S}_1 & : \text{covariance matrix of sample } \mathbf{X}, \\
\mathbf{S}_2 & : \text{covariance matrix of sample } \mathbf{Y}, \\
n_1 & : \text{number of observations in sample } \mathbf{X}, \\
n_2 & : \text{number of observations in sample } \mathbf{Y}, \\
p & : \text{number of variables (dimensions)}, \\
(\cdot)^\top & : \text{transpose operator}.
\end{align*}

The equations \ref{eq:hotelling_t2}, \ref{eq:mean_X}, \ref{eq:mean_Y}, \ref{eq:pooled_cov}, \ref{eq:sample_cov_X}, and \ref{eq:sample_cov_Y} shows the steps that how we calculate Hotelling's $T^{2}$ statistic, and we have named the variables for better understanding. 

In this study, the multivariate observations correspond to the pixel coordinates contained within the contours of each coronal hole (CH).  
Specifically, $\mathbf{X}$ represents contours predicted by the neural network (NN), while $\mathbf{Y}$ represents contours derived from the ground truth (GT).

\vspace{0.4cm}
\noindent\textbf{Null hypothesis ($H_0$):}  
The mean contour obtained from the NN does not differ from the mean contour obtained from the GT:
\begin{equation}
H_0 : \boldsymbol{\mu}_{\mathrm{NN}} = \boldsymbol{\mu}_{\mathrm{GT}}
\label{eq:null-hypothesis}
\end{equation}

\noindent\textbf{Alternative hypothesis ($H_1$):}  
The mean contour obtained from the NN differs from the mean contour obtained from the GT:
\begin{equation}
H_1 : \boldsymbol{\mu}_{\mathrm{NN}} \neq \boldsymbol{\mu}_{\mathrm{GT}}
\label{eq:alternative-hypothesis}
\end{equation}
\section{Model Input Features}
\label{sec:model_input_features}

The model inputs consist of a time-indexed dataset with 354,502 samples spanning from 
01 May 1996 to 16 July 2022. All input variables are binary-valued 
($0/1$) and encoded as integers. The complete feature set comprises 93 input variables, 
organized into thematic categories as described below.

\subsection{Active Region (AR) Location Categories}
\label{subsec:AR}
\begin{itemize}
  \item \texttt{cat\_AR\_CM+}, \texttt{cat\_AR\_CM-}
  \item \texttt{cat\_AR\_E+}, \texttt{cat\_AR\_E-}
  \item \texttt{cat\_AR\_SE+}, \texttt{cat\_AR\_SE-}
  \item \texttt{cat\_AR\_W+}, \texttt{cat\_AR\_W-}
  \item \texttt{cat\_AR\_C}
\end{itemize}

\subsection{Flare (FL) Location Categories}
\label{subsec:FL}
\begin{itemize}
  \item \texttt{cat\_FL\_CM+}, \texttt{cat\_FL\_CM-}
  \item \texttt{cat\_FL\_E+}, \texttt{cat\_FL\_E-}
  \item \texttt{cat\_FL\_W+}, \texttt{cat\_FL\_W-}
  \item \texttt{cat\_FL\_SE+}, \texttt{cat\_FL\_SE-}
  \item \texttt{cat\_FL\_C}
\end{itemize}

\subsection{Coronal Mass Ejection (CME) Location Categories}
\label{subsec:CME}
\begin{itemize}
  \item \texttt{cat\_CME\_CM+}, \texttt{cat\_CME\_CM-}
  \item \texttt{cat\_CME\_E+}, \texttt{cat\_CME\_E-}
  \item \texttt{cat\_CME\_SE+}, \texttt{cat\_CME\_SE-}
  \item \texttt{cat\_CME\_W+}, \texttt{cat\_CME\_W-}
  \item \texttt{cat\_CME\_C}
\end{itemize}

\subsection{GOES Flare Classification}
\label{subsec:GOES}
\begin{itemize}
  \item \texttt{fl\_goescls\_X}
  \item \texttt{fl\_goescls\_M}
  \item \texttt{fl\_goescls\_C}
  \item \texttt{fl\_goescls\_B}
  \item \texttt{fl\_goescls\_A}
\end{itemize}

\subsection{Solar Cycle Phase Indicators}
\label{subsec:SCPhase}
\begin{itemize}
  \item \texttt{phase\_declining}
  \item \texttt{phase\_rising}
  \item \texttt{phase\_solar\_max}
  \item \texttt{phase\_solar\_min}
\end{itemize}

\subsection{Mount Wilson Magnetic Classification}
\label{subsec:MountWilson}
\begin{itemize}
  \item \texttt{ar\_mtwilsoncls\_05TAA}
  \item \texttt{ar\_mtwilsoncls\_ALPHA}
  \item \texttt{ar\_mtwilsoncls\_ALPHADELTA}
  \item \texttt{ar\_mtwilsoncls\_ALPHADELTA-DELTA}
  \item \texttt{ar\_mtwilsoncls\_ALPHAGAMMA}
  \item \texttt{ar\_mtwilsoncls\_ALPHAGAMMA-DELTA}
  \item \texttt{ar\_mtwilsoncls\_BETA}
  \item \texttt{ar\_mtwilsoncls\_BETA-DELTA}
  \item \texttt{ar\_mtwilsoncls\_BETA-DELTA-DELTA}
  \item \texttt{ar\_mtwilsoncls\_BETA-GAMMA}
  \item \texttt{ar\_mtwilsoncls\_BETA-GAMMA-DELTA}
  \item \texttt{ar\_mtwilsoncls\_BETAA}
  \item \texttt{ar\_mtwilsoncls\_BETAADELTA}
  \item \texttt{ar\_mtwilsoncls\_BETAADELTA-DELTA}
  \item \texttt{ar\_mtwilsoncls\_BETAAGAMMA}
  \item \texttt{ar\_mtwilsoncls\_BETAAGAMMA-DELTA}
\end{itemize}

\subsection{Emerging Flux (EF) Location Categories}
\label{subsec:EF}
\begin{itemize}
  \item \texttt{cat\_EF\_E+}, \texttt{cat\_EF\_E-}
  \item \texttt{cat\_EF\_W+}, \texttt{cat\_EF\_W-}
  \item \texttt{cat\_EF\_CM+}, \texttt{cat\_EF\_CM-}
  \item \texttt{cat\_EF\_SE+}, \texttt{cat\_EF\_SE-}
  \item \texttt{cat\_EF\_C}
\end{itemize}

\subsection{Filament Eruption (FE) Location Categories}
\label{subsec:FE}
\begin{itemize}
  \item \texttt{cat\_FE\_CM+}, \texttt{cat\_FE\_CM-}
  \item \texttt{cat\_FE\_E+}, \texttt{cat\_FE\_E-}
  \item \texttt{cat\_FE\_SE+}, \texttt{cat\_FE\_SE-}
  \item \texttt{cat\_FE\_W+}, \texttt{cat\_FE\_W-}
  \item \texttt{cat\_FE\_C}
\end{itemize}

\subsection{Filament (FI) Location Categories}
\label{subsec:FI}
\begin{itemize}
  \item \texttt{cat\_FI\_CM+}, \texttt{cat\_FI\_CM-}
  \item \texttt{cat\_FI\_E+}, \texttt{cat\_FI\_E-}
  \item \texttt{cat\_FI\_SE+}, \texttt{cat\_FI\_SE-}
  \item \texttt{cat\_FI\_W+}, \texttt{cat\_FI\_W-}
  \item \texttt{cat\_FI\_C}
\end{itemize}

\subsection{Coronal Dimming (CD) Location Categories}
\label{subsec:CD}
\begin{figure}[htbp]
    \centering
    \includegraphics[width=0.35\textwidth]{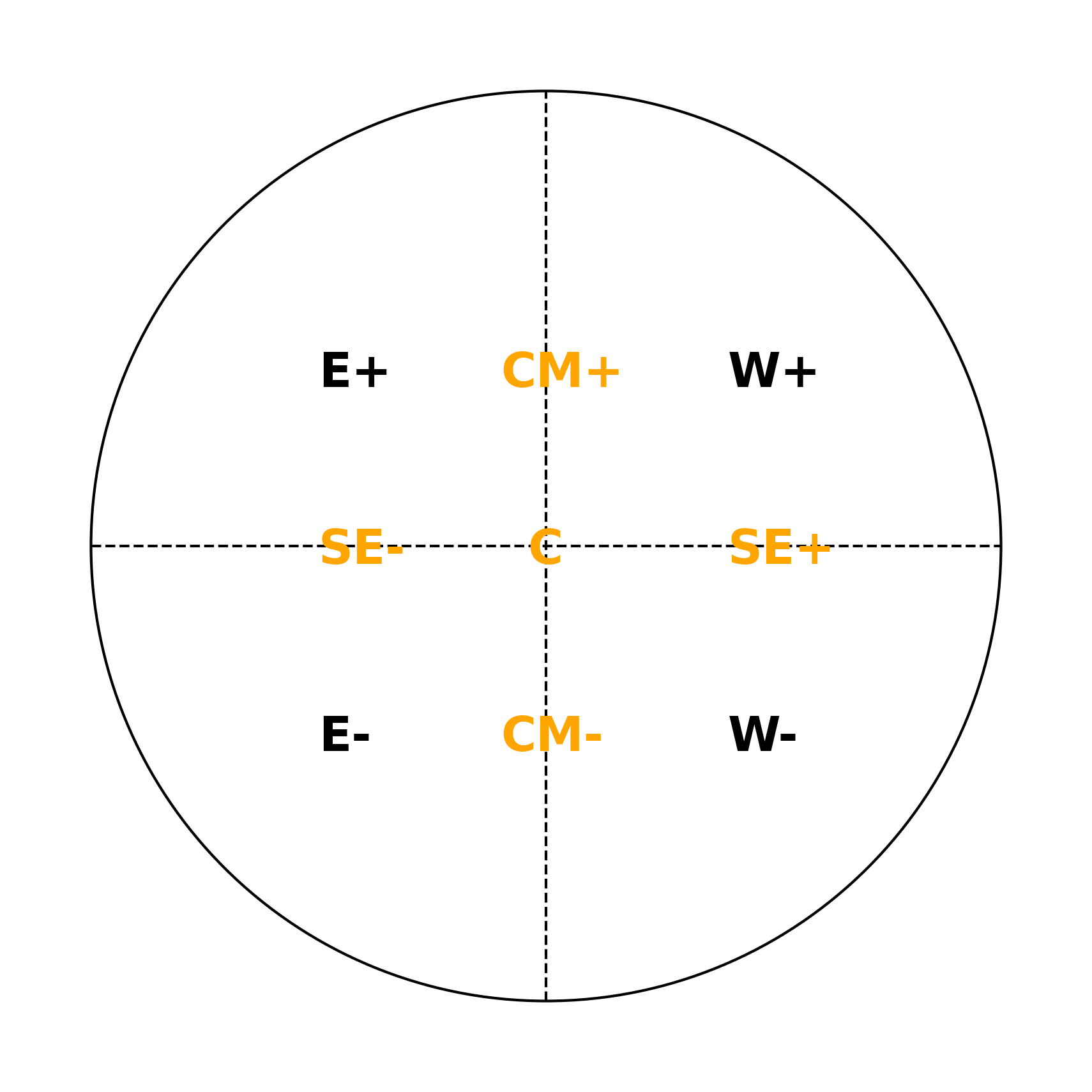}
    \caption{Quadrants of the solar disk to illustrate the spatial distributions for training.}
    \label{fig:quadrant}
\end{figure}

\begin{itemize}
  \item \texttt{cat\_CD\_CM+}, \texttt{cat\_CD\_CM-}
  \item \texttt{cat\_CD\_E+}, \texttt{cat\_CD\_E-}
  \item \texttt{cat\_CD\_W+}, \texttt{cat\_CD\_W-}
  \item \texttt{cat\_CD\_SE+}, \texttt{cat\_CD\_SE-}
  \item \texttt{cat\_CD\_C}
\end{itemize}
\noindent
Here, in Figure~\ref{fig:quadrant}, we represent the quadrants of the solar disk to illustrate the spatial distributions for training, and the meanings of the notations as below,
\begin{enumerate}
  \item Central meridian (\textbf{CM})
  \item Solar equator (\textbf{SE})
  \item Eastern limb (\textbf{E})
  \item Western limb (\textbf{W})
  \item Center (\textbf{C})
\end{enumerate}

\section{Relation between the threshold values in SDO \& SOHO}
\label{sec:appendix_SDO-SoHO}

We also perform Hotelling's $T^2$, test to validate the SDO-SoHO conversion for the date of \textit{2010-06-23 00:00:06.60} and from the statistics we use the same hypothesis from the Appendix~\ref{sec:HT}, but this time comparing the pixel coordinates within contours from SDO and SoHO. The $T^2$ = 3.53 and p-value = 0.17 indicate that the null hypothesis is not rejected, suggesting that the SDO image and the SoHO conversion for this specific date are statistically consistent. Likewise, we train the model for the SDO-SoHO conversion, and we plan to release the improved version on GitHub in the future.

\section{POP-CORN in GitHub}
\label{sec:Github}

Please refer to the GitHub page for POP-CORN through: \citep{Henadhira2026}.

Here, the user has the ability to download the CH contour coordinates and the pixel coordinates within the contours for the user-selected resolution.

\end{document}